\documentclass[twocolumn,trackchanges]{aastex631}

\usepackage{amsmath}
\usepackage{multirow}
\usepackage{cancel}

\begin{document}

\title{How many supermassive black hole binaries are detectable \\through tracking relative motions by sub/millimeter VLBI}

\correspondingauthor{Shan-Shan Zhao}
\email{zhaoss@shao.ac.cn}

\author[0000-0002-9774-3606]{Shan-Shan Zhao}
\affiliation{Shanghai Astronomical Observatory, Chinese Academy of Sciences, \\80 Nandan Road, Shanghai 200030, People’s Republic of China}

\author[0000-0001-7369-3539]{Wu Jiang}
\affiliation{Shanghai Astronomical Observatory, Chinese Academy of Sciences, \\80 Nandan Road, Shanghai 200030, People’s Republic of China}
\affiliation{Key Laboratory of Radio Astronomy and Technology, Chinese Academy of Sciences, A20 Datun Road, Chaoyang District, Beijing 100101, People's Republic of China}

\author[0000-0002-7692-7967]{Ru-Sen Lu}
\affiliation{Shanghai Astronomical Observatory, Chinese Academy of Sciences, \\80 Nandan Road, Shanghai 200030, People’s Republic of China}
\affiliation{Key Laboratory of Radio Astronomy and Technology, Chinese Academy of Sciences, A20 Datun Road, Chaoyang District, Beijing 100101, People's Republic of China}
\affiliation{Max-Planck-Institut für Radioastronomie, Auf dem Hügel 69, D-53121 Bonn, Germany}

\author[0000-0002-1923-227X]{Lei Huang}
\affiliation{Shanghai Astronomical Observatory, Chinese Academy of Sciences, \\80 Nandan Road, Shanghai 200030, People’s Republic of China}

\author[0000-0003-3540-8746]{Zhiqiang Shen}
\affiliation{Shanghai Astronomical Observatory, Chinese Academy of Sciences, \\80 Nandan Road, Shanghai 200030, People’s Republic of China}
\affiliation{Key Laboratory of Radio Astronomy and Technology, Chinese Academy of Sciences, A20 Datun Road, Chaoyang District, Beijing 100101, People's Republic of China}



\begin{abstract}
The sub/millimeter wavelengths (86-690 GHz) very long baseline interferometry (VLBI) will provide $\sim5-40\ \mu$as angular resolution, $\sim10$ mJy baseline sensitivity, and $\sim 1\ \mu$as/yr proper motion precision, which can directly detect supermassive black hole binary (SMBHB) systems by imaging the two visible sources and tracking their relative motions. Such a way exhibits an advantage compared to indirect detect methods of observing periodic signals in motion and light curves, which are difficult to confirm from competing models. Moreover, tracking relative motion at sub/millimeter wavelengths is more reliable, as there is a negligible offset between the emission region and the black hole center. In this way, it is unnecessary to correct the black hole location by a prior of jet morphology as it would be required at longer wavelengths. We extend the formalism developed in \cite{DOrazio2018ApJ863.185} to link the observations with the orbital evolution of SMBHBs from the $\lesssim$10 kpc dynamical friction stages to the $\lesssim 0.01$ pc gravitational radiation stages, and estimate the detectable numbers of SMBHBs. By assuming 5\% of AGNs holding SMBHBs, we find that the number of detectable SMBHBs with redshift $z\le 0.5$ and mass $M\leq 10^{11}M_\sun$ is about 20. Such detection relies heavily on proper motion precision and sensitivity. Furthermore, we propose that the simultaneous multi-frequency technique plays a key role in meeting the observational requirements.
\end{abstract}



\section{Introduction} \label{sec:intro}

Using the very long baseline interferometry (VLBI) techniques, the Event horizon telescope (EHT) has achieved direct imaging of supermassive black holes (SMBHs) in the center of M87 and our Galaxy \citep{EHTC2019ApJL875.L1,EHTC2022ApJ930L12}. The high resolution of EHT ($\sim 20\ \mu$as) is provided by the short observing wavelengths (1.3 mm, i.e., 230 GHz) and the long baselines (10,700 km). Sub/millimeter (refers to 86-690 GHz in this work) VLBI is expected to directly observe supermassive black hole binaries (SMBHBs) in galaxies by tracking the relative motion of the two visible sources. Inspired by \cite{Pesce2021ApJ923.260} about the estimate of the number of observable black hole shadows by the next generation EHT (ngEHT), we aim to answer the question of how many SMBHBs can be detected by the sub/millimeter VLBI. 

It is generally believed that SMBHBs are formed from the merger of their host galaxies. Different mechanisms drive the black holes to move towards each other from the galaxy scale ($\lesssim$ 10 kpc) to the sub-parsec scale ($\lesssim$ 0.01 pc), and finally merge. On the theoretical aspect, it is relatively clear that the driven mechanism at the large scales ($\sim$10 kpc to $\sim$10 pc) is the dynamical friction\footnote{The source holding SMBHB at this stage is called dual AGN.} \citep[][]{Chandrasekhar1943ApJ97.255}, while on the small scales ($\sim$0.01 pc to mergers) binary systems lose the angular momentum due to emitting gravitational waves \citep[GWs,][]{Peters1964PhRv136.1224}. However, on the intermediate scales, there is a dispute over the driving mechanisms, i.e., the so-called "final parsec problem" \citep[e.g.,][]{Begelman1980Natur287.307,Binney2008GalacticDynamics,Merritt2013CQGra30.244005,Vasiliev2015ApJ810.49}. Gaseous environments \citep{Munoz2020ApJ889.114,Dittmann2022MNRAS513.6158,Lai2022arXiv2211.00028}, star scattering \citep{Merritt2005LRR8.8}, and dark matter\citep{Kelley2017MNRAS464.3131} may contribute to solve the problem. The physical plausibility of those ideas has been widely explored through simulations \citep[e.g.,][]{Kelley2017MNRAS464.3131,Kelley2019MNRAS485.1579,Munoz2020ApJ889.114,Chen2022MNRAS510.531,Chen2022MNRAS514.2220,Volonteri2022MNRAS514.640}, and requires further tests through observations.

The indirect observations, such as analyzing periodic behaviors in the light curves \citep[e.g.,][]{Burke-Spolaor2018ASPC517.677} or the motions \citep[e.g.,][]{Sudou2003Sci300.1263}, as well as resolving double-peaked emission lines \citep[e.g.,][]{An2018RaSc53.1211,Breiding2021ApJ914.37,Saade2020ApJ900.148}, require excluding competing models and independent confirmations. Direct observations are more challenging. One direct observation is to detect the low frequency GWs by Pulsar Timing Arrays \citep[PTAs, e.g.,][]{Mingarelli2017NatAs1.886,DOrazio2018ApJ863.185,Arzoumanian2021ApJ914.121}.  The other way is to directly track binaries' relative motions with VLBI, which is what this work focuses on. It requires that both sources are visible and their separation and relative motions are measurable. \cite{Bansal2017ApJ843.14} successfully measured $1.57\ \mu$as/yr relative motion of SMBHB 0402+379 by 12 years of VLBI observations at 5-22 GHz. However, due to the existence of core-shift \citep[the position of the radio core moves towards the central engine with increasing frequency, e.g. M87 case by][]{Hada2011Natur477.185}, it needs a precise jet morphology model to locate the centre black hole from the observed jet core at low frequencies. At high frequencies, however, the radio core can be approximated to be located at the center of the black hole without any prior knowledge of jet morphology to determine its location. As a result, tracking relative motion through sub/millimeter VLBI provides a method for detecting SMBHBs that is both observationally direct and physically clean.

In this paper, we follow the formalism from \cite{DOrazio2018ApJ863.185} to predict the population of observable SMBHBs using an updated prescription for the binary orbital evolution model. The prediction is for all observable AGNs with redshift $z\le 0.5$ and mass $M\lesssim 10^{11}M_\sun$. The result is constrained by minimum detectable angular velocity, minimum detectable angular separation and minimum detectable flux density. Next, taking into account the possible capabilities of proper motion precision, angular resolution, and baseline sensitivity, we discuss the feasibility of observing SMBHBs at 86, 230, 345, and 690 GHz via VLBI.

This paper is written using the following structure. In Section~\ref{sec:resident_time}, we introduce the model of binary orbital evolution from which we calculate the resident timescale of SMBHBs. In Section~\ref{sec:number_SMBHBs}, we show our method to compute the number of detectable SMBHBs by combining the probability distribution function (PDF) with the mass-luminosity model and the radio luminosity functions (RLFs). The results at multi-frequencies are displayed and analyzed in Section~\ref{sec:results}. In Section~\ref{sec:techniques}, we show how to achieve the observational requirements. The conclusions and discussions are in Section~\ref{sec:conclusion}. 

The cosmology used in this paper is $H_0 = 71$ km s$^{-1}$ Mpc$^{-1}$, $\Omega_{M}$=0.27, $\Omega_{\Lambda}$=0.73.

\section{Binary orbital evolution}
\label{sec:resident_time}

Suppose a satellite galaxy merges in a host galaxy. The host galaxy contains a supermassive black hole (SMBH) of mass $M_1$, and the satellite galaxy contains a secondary SMBH of mass $M_2$, where 
\begin{equation}
    M_1=\frac{1}{1+q}M,\ M_2=\frac{q}{1+q}M,
\end{equation}
$M=M_1+M_2$ is the total mass of the system, $q$ is called the mass ratio and $q\leq1$.
The separation between two black hole $a$ shrinks from the initial separation $a_{\rm ini}=10$ kpc to zero (merger). Such separation shrinkage can be described by a model of binary orbital evolution, which is caused by several different driven mechanisms. Following \cite{Begelman1980Natur287.307}, our model contains four parts: evolution from dynamical friction, evolution from stellar hardening, evolution from gas accretion, and evolution from gravitational waves. In this model, we assume the host galaxy is rich of gas at the central region, so that the loss cone depletion phenomena can be neglected. We focus on two quantities changing with orbital evolution: the resident timescale $t_{\rm res}=|a/\dot{a}|$  and the relative velocity between the two black holes.

In the following subsection, we will demonstrate each phase of the orbital evolution model (Section~\ref{sec:dynamical_friction} to \ref{sec:gravitational_radiation}) and summarize them in Section~\ref{sec:resident_time_summary}.

\subsection{dynamical friction}
\label{sec:dynamical_friction}

When the separation shrinks from $a_{\rm ini}$ to $a_h$ (hardening radius, defined by Eq.~\eqref{eq:a_h} in Sect.~\ref{sec:hardening}), the secondary black hole moves from the edge of the host galaxy to the center region and travels through the field of stars in the host galaxy. The energy is transferred from the relative orbital motion of the secondary black hole to the random motion of the stars and thus the orbit decays. Such effects can be approximately described by the dynamical friction, which can be expressed by the Chandrasekhar formalism \citep{Chandrasekhar1943ApJ97.255}. A simplified application of Chandrasekhar’s formula \citep[see ][]{Binney2008GalacticDynamics} can be adopted with assuming: i) the density distribution of the host galaxy can be approximated by a singular isothermal sphere; ii) the host galaxy's velocity distribution is Maxwellian; iii) the secondary black hole moves in a circular orbit at a velocity $v_{\rm DF}$ relative to the host black hole, which is 
\begin{equation}
    v_{\rm DF}=\sqrt{2}\sigma_1,
\end{equation}
where $\sigma_1$ is the velocity dispersion of the host galaxy. The dynamical friction timescale $t_{\rm DF}$ under the above assumptions is deduced by
\cite{Binney2008GalacticDynamics}. It is then modified by \cite{Dosopoulou2017ApJ840.31}, who assume that a mass $1000 M_2$ of stars are bound to the secondary black hole. This setup yields a corresponding dynamical friction timescale of
\begin{equation}
\begin{split}
     t_{\rm DF}&= 
    0.12\ {\rm Gyr}\\
    &\times\left(\!\frac{a_{\rm ini}}{10\ {\rm kpc}}\!\right)^{\!2}\!\left(\!\frac{\sigma_1}{300\ {\rm km\ s^{\!-1}}}\!\right)\!\left(\!\frac{M_2}{10^8 M_\odot}\!\right)^{\!-1}\!\! \frac{1}{\ln{\Lambda}},  \label{eq:tres_DF}    
\end{split}
\end{equation}
where $\ln{\Lambda}$ is the Coulomb logarithm:
\begin{equation}
    \Lambda=\frac{b_{\rm max}}{(GM_2)/v^2_{\rm DF}},\ b_{\rm max}=10\ {\rm kpc},
    \label{eq:Lambda}
\end{equation}
$G$ is the gravitational constant. The timescale is supported by the simulation results of \cite{Chen2022MNRAS510.531,Chen2022MNRAS514.2220}. During the dynamical friction stage, the orbital separation is decreasing exponentially with an e-folding timescale of $t_{\rm DF}$.

Here we do not consider the rotation of the secondary galaxy and the non-isothermal structures such as bars and spiral arms in the host galaxy. We also do not consider the different initial velocity of the secondary black hole, which may lead to a non-circular orbit dragged by the dynamical friction force. See \cite{Binney2008GalacticDynamics} for more complex cases. 

We calculate $\sigma_1$ from the $M_{\rm BH}-\sigma$ relation given by \cite{Tremaine2002ApJ574.740}: 
\begin{equation}
    \log_{10}\left(\frac{M_1}{M_\odot}\right)=8.13+4.02\log_{10}\left(\frac{\sigma_1}{200\ {\rm km\ s^{-1}}}\right),
    \label{eq:M_sigma}
\end{equation}
which is consistent with the simulation by \cite{Chen2022MNRAS514.2220}. 

\subsection{stellar hardening}
\label{sec:hardening}
When the separation $a$ decreases to $a\ll GM/\sigma_1^2$, the binaries tend to be in the hard state. In this state, the relative velocity for a circular orbit is \citep{Binney2008GalacticDynamics}:
\begin{equation}
\label{eq:v_h}
    v_h=\sqrt{\frac{GM}{a}}.
\end{equation}
Chandrasekhar’s dynamical friction formula is no longer valid and the individual interactions between each star and the binary system should be considered. We define the hardening radius $a_h$ as the boundary between the dynamical friction process and the hardening state. Following \cite{Begelman1980Natur287.307}, $a_h$ is
\begin{equation}
    \label{eq:a_h}
  a_h\equiv\left(\frac{M_1}{M_{\rm 1, core}}\right)r_{\rm core}= \frac{GM_1}{3\sigma_1^2},
\end{equation}
 where we use the King radius $r_0\equiv \left[(9\sigma^2)/(4\pi G\rho_0)\right]^{1/2}$ as the core radius and assume that the core is spherical and described by a constant density $\rho_0$ \citep[for more details, see][]{Binney2008GalacticDynamics}. The hardening rate can be approximately expressed by a similar formula from the hardening rate of stellar binaries given by \cite{Quinlan1996NewA1.35}:
\begin{equation}
    \frac{d}{dt}\left(\frac{1}{a}\right)=-C\frac{G\rho_0}{\sigma_1},
\end{equation}
where $C$ is the hardening rate coefficient. Following \cite{Binney2008GalacticDynamics}, we adopt $C=14.3$. We use $r_0$ to replace $\rho_0$, then the resident timescale is
\begin{equation}
        t_h=\frac{4\pi r_0^2}{9C\sigma_1 a}.
        \label{eq:tres_h}
\end{equation}
We choose $r_0=100$ pc for a typical core size. 

Here we simply let the binary enter the hardening state directly when $a\to a_h$, and it generates discontinuity in evolution of velocity and resident timescale. As far as computational precision is concerned, such a simplification is tolerable. See \cite{Begelman1980Natur287.307} for a more complicated hardening process. It starts with the secondary black hole reaching the core region, then the binary is bound up by gravity and finally becomes hard.

\subsection{gaseous accretion}
When $a$ decreases to $a_{\rm gas}$, there are no stars in the region and gaseous environment plays the key role in the orbital evolution. The velocity $v_{\rm gas}$ of gaseous state is described by the same formula as $v_h$ given by Eq.~\eqref{eq:v_h}. 
The timescale of gaseous state is \citep{DOrazio2018ApJ863.185}:
\begin{equation}
        t_{\rm gas} = \frac{q_s}{4\dot{\mathcal{M}}} t_{\rm Edd},
        \label{eq:tres_gas}
\end{equation}
where $q_s\equiv4q/(1+q)^2$ is the symmetric binary mass ratio, $\dot{\mathcal{M}}=\dot{M}/\dot{M}_{\rm Edd}$ is the Eddington rate and $t_{\rm Edd}\equiv M/\dot{M}_{\rm Edd}$ is the Eddington time. We set $t_{\rm Edd}\sim 4.5\times 10^7$ yr (assuming accretion efficiency is $\sim 0.1$) and $\dot{\mathcal{M}}\sim 1$ to be consistent with the previous work \citep{DOrazio2018ApJ863.185}. 
We assume $a_{\rm gas}$ is where $t_h=t_{\rm gas}$:
\begin{equation}
    a_{\rm gas}=\frac{16\pi r_0^2\dot{\mathcal{M}}}{9C t_{\rm Edd} \sigma_1 q_s}.   
    \label{eq:a_gas}
\end{equation}

Notice, in this stage, we do not consider the loss-cone depletion \citep[e.g.,][]{Merritt2013CQGra30.244005}, which may exist if the environment of the binary system is poor of gas. The loss-cone regime is where the binary loses angular momentum via scattering the stars. When the binary empties the loss-cone, the orbital shrinkage stalls and it will take a very long time to refill the loss-cone. 

\subsection{gravitational radiation}
\label{sec:gravitational_radiation}

When $a$ is smaller than $a_{\rm GW}$, the binary is close enough that the system keeps losing angular momentum through emitting GWs until they merge. The resident timescale of evolution from gravitational radiation can be approximated to the lifetime of a binary system from separation $a$ ($a\leq a_{\rm GW}$) to merger, which is determined by \cite{Peters1964PhRv136.1224}:
\begin{equation}
        t_{\rm GW}=\frac{5}{64}\frac{c^5a^4}{G^3M^3}q_s^{-1},
        \label{eq:tres_GW}
\end{equation}
and it begins when the orbital separation reaches
\begin{equation}
    a_{\rm GW}=2\sqrt{q_s}\left[\frac{t_{\rm Edd}}{5 c^5 \dot{\mathcal{M}}}(GM)^3\right]^{1/4},
\end{equation}
where $c$ is the speed of light. $a_{\rm GW}$ is where $t_{\rm gas}=t_{\rm GW}$.

\subsection{summary}
\label{sec:resident_time_summary}

Our binary orbital evolution model assumes that all SMBHBs start out in a circular orbit with an initial orbital separation of $a_{\rm ini} = 10$ kpc. They have an initial orbital velocity determined by the stellar velocity dispersion in the host galaxy. During each phase of orbital evolution, we assume that the orbit shrinks according to the differential equation $|a / \dot{a}| = t_{\rm res}$, with $t_{\rm res}$ being the resident timescale for that phase. 
\begin{itemize}
    \item At the largest separations, the binary evolves through dynamical friction, during which the orbit shrinks exponentially with an e-folding timescale of $t_{\rm res}$ (see Eq.~\eqref{eq:tres_DF}).
    \item Once the binary has shrunk to an orbital separation of $a = a_{\rm h} \approx $ several pc (see Eq.~\eqref{eq:a_h}), we assume that the evolution becomes instantaneously dominated by stellar hardening, during which the orbital separation evolves proportionally to the reciprocal of the time and the residence time increases as $t_{\rm res} \propto a^{-1}$ (see Eq.~\eqref{eq:tres_h}).
    \item Once the residence time has increased with shrinking $a$ to $t_{\rm res} = t_{\rm gas}$ (see Eq.~\eqref{eq:a_gas}), we assume that the evolution becomes instantaneously dominated by gaseous accretion, during which the orbit once again shrinks exponentially with an e-folding timescale of $t_{\rm res}$.
    \item Finally, once the binary has shrunk to the point where the resident time is equal to $t_{\rm GW}$ (see Eq.~\eqref{eq:tres_GW}), we assume that the evolution becomes instantaneously dominated by gravitational wave emission, after which the orbit decays in a very short time (given by Eq.~\eqref{eq:tres_GW}). 
\end{itemize}

\begin{figure*}
\centering
    \includegraphics[scale=0.95]{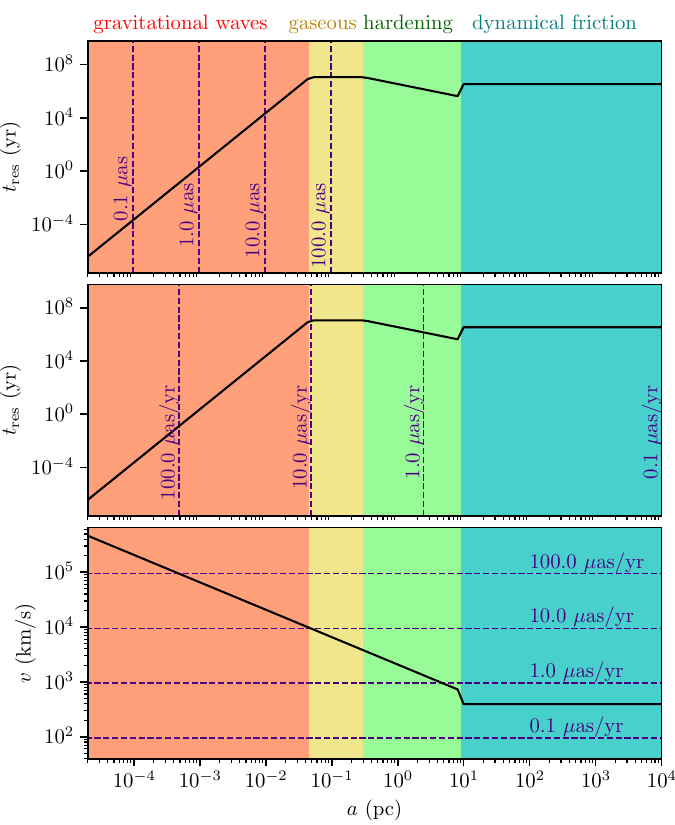}
    \includegraphics[scale=0.95]{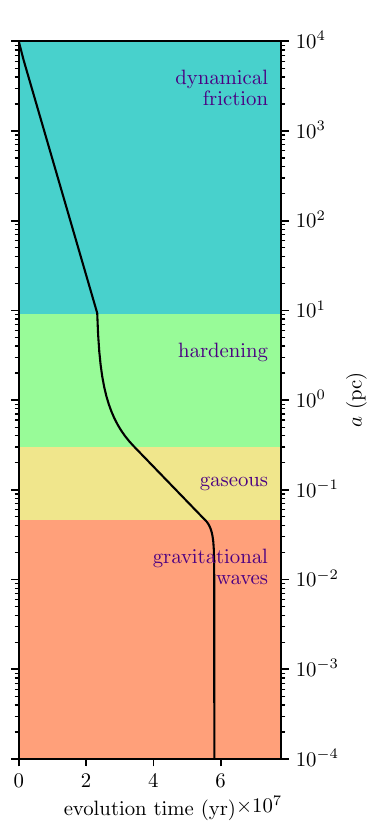}
    \caption{Left panel: The resident timescale (top and middle panel) and the velocity (bottom panel) of orbital evolution of a SMBHB system with $M=10^9\ M_\odot$, $q=1$ (solid lines). The colored backgrounds demonstrate the four stages of orbital evolution: dynamical friction (blue), hardening (green), gaseous state (yellow) and GWs state (orange). The dashed lines present the corresponding $a$ and $v$ when the observed angular separation of the two sources is 0.1, 1, 10, 100 $\mu$as and their relative motion is 0.1, 1, 10, 100 $\mu$as/yr, assuming $z=0.05$. Right panel: This plot shows the orbital shrinkage from $a_{\rm ini}=10$ kpc to 0, with an evolution time of 5.8$\times 10^7$ yr, during the four phases. It corresponds to the case shown in the left panel.}
    \label{fig:orbital_decay}
\end{figure*}

The various ranges of orbital separation, orbital velocity, and resident timescales associated with these four evolutionary phases are summarized in Tab.~\ref{tab:resident_time}, and the evolution of an example SMBHB system with $M=10^9\ M_\odot$, $q=1$ is illustrated in Fig.~\ref{fig:orbital_decay}.

\begin{table*}[]

    \centering
    \caption{The summary of orbital evolution of SMBHBs.}
    \renewcommand{\arraystretch}{1.8}

     \begin{tabular}{l|l|l|l}
        \hline
        \hline
         driven mechanisms & $a$  & $v(M,q,a)$& $t_{\rm res}(M,q,a)$ \\
        \hline
        dynamical friction & from $a_{\rm ini}$ to $a_h$ & $\sqrt{2}\sigma_1$ & $1.2\times10^8\ {\rm yr}\left(\frac{a_{\rm ini}}{10\ {\rm kpc}}\right)^2\left(\frac{\sigma_1}{300\ {\rm km\ s^{-1}}}\right)\left(\frac{M_2}{10^8 M_\odot}\right)^{-1} \ln^{-1}\Lambda $\\
        hardening & from $a_h$ to $a_{\rm gas}$ & $\left(\frac{GM}{a}\right)^{1/2}$& $3.2\times10^6\ {\rm yr}\left(\frac{\sigma_1}{300\ {\rm km\ s^{-1}}}\right)^{-1}\left(\frac{a}{1\ {\rm pc}}\right)^{-1}$\\
        gaseous decay & from $a_{\rm gas}$ to $a_{\rm GW}$  & $\left(\frac{GM}{a}\right)^{1/2}$ & $1.1\times10^7\ {\rm yr}\left(\frac{1.0}{\dot{\mathcal{M}}}\right)\ q_s$\\
        gravitational waves & from $a_{\rm GW}$ to 0 &  $\left(\frac{GM}{a}\right)^{1/2}$ &$85\ {\rm day} \left(\frac{a}{10^{-4} \ {\rm pc}}\right)^{4}\left(\frac{M}{10^8 M_\odot}\right)^{-3}q_s^{-1}$\\[1em]
        \hline
    \end{tabular}

    \label{tab:resident_time}
\end{table*}

\section{Method of estimating the number of detectable SMBHBs}
\label{sec:number_SMBHBs}

The number of detectable SMBHBs $N_\nu(\dot{\theta}_{\rm min},\theta_{\rm min},F_{\rm min})$ at observing frequency $\nu$ is constrained by three parameters: minimum detectable angular velocity $\dot{\theta}_{\rm min}$, minimum detectable angular separation $\theta_{\rm min}$ and minimum detectable flux density $F_{\rm min}$. It can be described by the integral of the distributions of detectable cases with redshift from 0 to $z_{\rm max}$ and luminosity from $L_\nu^{\rm min}(F_{\rm min})$ to infinity, which is \citep{DOrazio2018ApJ863.185}  
\begin{equation}
\begin{split}
    N_\nu(\dot{\theta}_{\rm min},&\theta_{\rm min},F_{\rm min})\\
    =&4\pi\int_0^{z_{\rm max}} \int^\infty_{L_\nu^{\rm min}(F_{\rm min})}\Phi(L_{\nu},z)\\
    &\times\mathcal{F}\left[\dot{\theta}_{\rm min},\theta_{\rm min};M(L_\nu),z\right]\frac{dV}{dz} dL_\nu dz.
    \end{split}
    \label{eq:N_SMBHBs}
\end{equation}
$\Phi(L_{\nu},z)=dN_{\rm AGN}/(dL_\nu dV)$ is the RLF, which defines the number distribution of AGNs over luminosity $L_\nu$ and redshift $z$, see Sect.~\ref{sec:RLF}. $\mathcal{F}\left[\dot{\theta}_{\rm min},\theta_{\rm min};M(L_\nu),z\right]$ is the PDF of a AGN source containing a detectable SMBHB system (see Sect.~\ref{sec:PDF}), and in which the mass of the source $M(L_\nu)$ is determined by the mass-luminosity model (see Sect.~\ref{sec:luminosity}).  $V$ is the comoving volume and $dV$ in solid angle $d\Omega$ and redshift interval $dz$ is
\begin{equation}
    dV=D_H\frac{(1+z)^2D^2_A(z)}{E(z)}dzd\Omega,
\end{equation}
where $D_H\equiv c/H_0$, $D_A(z)$ is the angular diameter distance
\begin{equation}
    D_A(z)=\frac{D_H}{1+z}\int_0^z\frac{d z'}{E(z')},
\end{equation}
and
\begin{equation}
   E(z)\equiv\sqrt{\Omega_M(1+z)^3+\Omega_\Lambda}.
\end{equation}
$ L_\nu^{\rm min}(F_{\rm min})$ in Eq.~\eqref{eq:N_SMBHBs} is the lower limit of $L_\nu$, which means the minimum detectable luminosity. It is related to $F_{\rm min}$ by
\begin{equation}
    L_\nu^{\rm min}(F_{\rm min})=4\pi D^2_L F_{\rm min}(1+z)^{-\alpha-1},
    \label{eq:Lmin}
\end{equation}
where $D_L=(1+z)^2D_A$ is the luminosity distance. $\alpha$ is the spectral index, which is used to define the spectra of the source: $L_\nu\propto\nu^\alpha$. We adopt $\alpha=-0.5$ in the whole paper. By taking this value, the predict number of bright AGNs by our model is well consistent to observations, see Sect.~\ref{sec:RLF}. For the outer integral of the variable $z$ in Eq.~\eqref{eq:N_SMBHBs}, we choose the upper limit to be $z_{\rm max}=0.5$. Systems with z out of 0.5 is beyond the observe capability of millimeter VLBI, as suggested in \cite{DOrazio2018ApJ863.185}.

\subsection{probability distribution function}
\label{sec:PDF}
The PDF of the detectable SMBHBs is defined by \citep{DOrazio2018ApJ863.185}
\begin{equation}
\begin{split}
     \mathcal{F}&\left(\dot{\theta}_{\rm min},\theta_{\rm min};M,z \right)\\   
     &\ =f_{\rm bin}\frac{  \displaystyle{\int}_{q_1}^1\int_{P_{\rm min}(\theta_{\rm min})}^{P_{\rm max}(\dot{\theta}_{\rm min})}t_{\rm res}(M,q,P)\ dP\  dq  }{\displaystyle{\int}_{q_0}^1\int_{0}^{P_{\rm ini}}t_{\rm res}(M,q,P)\ dP\  dq},
\end{split}
\label{eq:PDF}
\end{equation}
where $f_{\rm bin}$ is the fraction of AGNs harboring SMBHBs. \cite{DOrazio2018ApJ863.185} suggested $f_{\rm bin}=0.05$ by taking account the consistency of the Pulsar Timing Arrays observations and the theoretical predictions of GWs backgrounds.
$t_{\rm res}(M,q,P)$ is the resident timescale of a SMBHB system with mass $M$, mass ratio $q$ and orbital period $P$. It can be deduced from $t_{\rm res}(M,q,a)$ (see Sect.~\ref{sec:resident_time}) by using the relation of $P$ and $a$:
\begin{equation}
    P=\frac{2\pi a}{v(M,q,a)},
\end{equation}
where $v(M,q,a)$ is summarized in Tab.~\ref{tab:resident_time}. 

In Eq.~\eqref{eq:PDF}, the numerator term is the integral of $t_{\rm res}$ over all the detectable $q$ and $P$ and the denominator term is the integral of $t_{\rm res}$ over all the physically possible $q$ and $P$. We assume that values of $q$ between 0.01 to 1 are physically possible as well as detectable, i.e. $q_0=q_1=0.01$. $P$ is physically possible from 0 to $P_{\rm ini}$, where
\begin{equation}
    P_{\rm ini}=\frac{2\pi a_{\rm ini}}{\sqrt{2}\sigma_1}.
\end{equation}
The upper limit of detectable $P$ is related to the minimum detectable angular velocity $\dot\theta_{\rm min}$, which is
\begin{equation}
    P_{\rm max}(\dot\theta_{\rm min})=\frac{2\pi a(M,q,{v_{\rm min})}}{v_{\rm min}},\ v_{\rm min}=\dot\theta_{\rm min} D_A(z).
    \label{eq:Pmax}
\end{equation}
If $P>P_{\rm max}$, the binary is orbiting too slowly to distinguish its relative motion. The lower limit of detectable $P$ is related to the minimum angular separation $\theta_{\rm min}$, which is
\begin{equation}
    P_{\rm min}(\theta_{\rm min})=\frac{2\pi a_{\rm min}}{v(M,q,a_{\rm min})},\ a_{\rm min}=\theta_{\rm min} D_A(z).
    \label{eq:Pmin}
\end{equation}
If $P<P_{\rm min}$, the binary is too close to be separated as two sources. 

Additionally, we assume $M$ is not likely to be larger than $M_{\rm max}=10^{11}M_\odot$, so we multiply a mass cut-off function $\exp{[-(M/M_{\rm max})^4}]$ on the PDF, as done by \cite{DOrazio2018ApJ863.185}.

\subsection{mass-luminosity model}
\label{sec:luminosity}

\begin{figure*}
    \centering
    \includegraphics[scale=0.8]{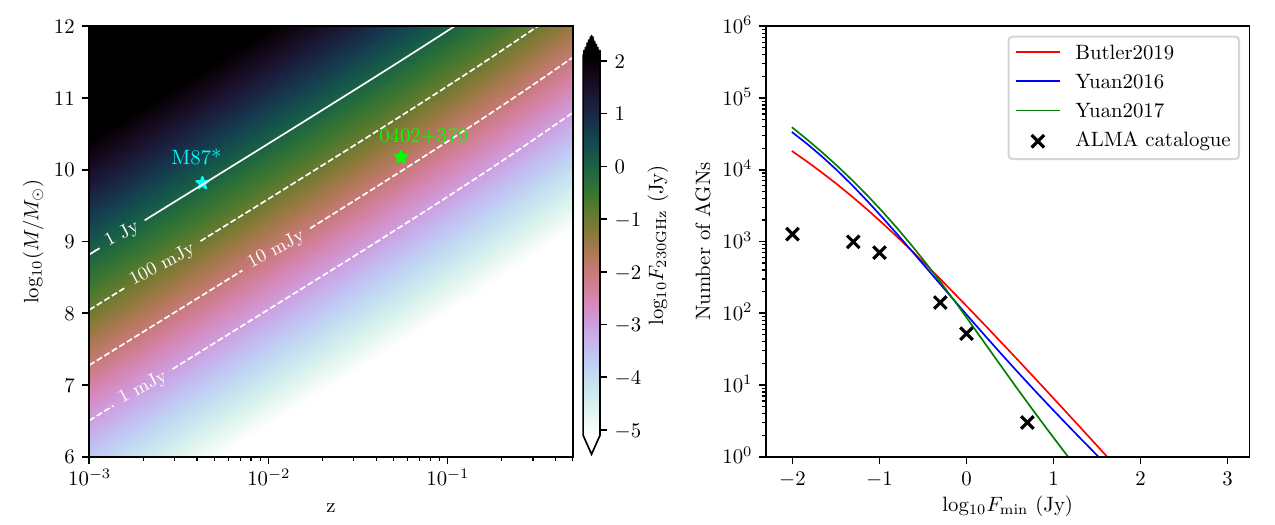}
    \caption{Left: The predicted 230 GHz flux density distribution of AGNs over mass M and redshift z, based on the mass-luminosity model Eq.~\eqref{eq:luminosity}. The cases of M87* ($M=6.5\times 10^9\ M_\odot$, z=0.004283) and SMBHB 0402+379 ($M=1.5\times 10^{10}\ M_\odot$, z=0.055) are marked with cyan and lime stars. Right: The predicted number of AGNs at 230 GHz on the entire sky with $z\leqslant0.5$, based on the RLFs estimated by \citet[][red line]{Butler2019AA625.111}, \citet[][blue line]{Yuan2016ApJ820.65}, \citet[][green line]{Yuan2017ApJ846.78}. It shows good consistency between the models. All predictions are reasonable when compared with the count of observed AGNs with a flux density larger than $F_{\rm min}$ from ALMA catalogue\textsuperscript{a} (black cross markers). }
    \label{fig:appendix}
    \begin{flushleft}
    \small\textsuperscript{a}ALMA Calibrator Source Catalogue: \url{https://almascience.eso.org/sc/}
    \end{flushleft}
\end{figure*}

We use the mass-luminosity model to estimate the mass of a AGN with luminosity $L_\nu$, which is
\begin{equation}
\begin{split}
     \log_{10}\left(\frac{M}{M_\odot}\right)=&\xi_{\rm R}\left[\log_{10}(\nu L_\nu)-(\alpha+1)\log_{10}\left(\frac{\nu}{\nu_0}\right)\right]\\
     &+\xi_{\rm X}\log_{10}\left(\frac{L_{\rm X}}{L_{\rm Edd}}\right)+\xi_{\rm C},
\end{split}
    \label{eq:luminosity_def}
\end{equation}
where $\xi_{\rm R}$, $\xi_{\rm X}$, $\xi_{\rm C}$ are coefficients, $\nu_0$ is the reference frequency. $L_{\rm X}$ is the X-ray luminosity
\footnote{The units of $L_\nu$ and $L_{\rm X}$ are erg s$^{-1}$ Hz$^{-1}$ and erg s$^{-1}$ in Eq.~\eqref{eq:luminosity_def}, while out of section \ref{sec:luminosity}, the unit of $L_\nu$ is W Hz$^{-1}$.},
$L_{\rm Edd}$ is the Eddington luminosity, which yields
\begin{equation}
    L_{\rm Edd}=\frac{4\pi G M m_p c}{\sigma_T}\simeq1.26\times10^{38}\left(\!\frac{M}{M_\odot}\!\!\right)\ {\rm erg \ s^{\!-1}}.
\end{equation}
By adopting the Fundamental Plane activities relation estimated at $\nu_0=$5 GHz by \cite{Plotkin2012MNRAS419.267}, we deduce the coefficients are: $\xi_{\rm R}=0.7713$, $\xi_{\rm X}=-0.5319$ and $\xi_{\rm C}=-23.49$. 
It is expected that $\log_{10}(L_{\rm X}/L_{\rm Edd})$ should vary between different types of AGNs. To simplify the model, we set it to a fixed value of -4.4 to obtain a good prediction of the relationship between the 230 GHz luminosity and the mass of M87* and SMBHB 0402+379. Then the mass-luminosity becomes,
\begin{equation}
\begin{split}
     \log_{10}\left(\frac{M}{M_\odot}\right)=&0.7713\log_{10}\left(\frac{L_\nu}{\rm erg\ s^{-1}\ Hz^{-1}}\right)\\
     -&0.7713\alpha\log_{10}\left(\frac{\nu}{5\rm GHz}\right)-13.67.
\end{split}
    \label{eq:luminosity}
\end{equation}
The left panel in Fig.~\ref{fig:appendix} shows the 
230 GHz flux density distribution of $M$ and $z$ predicted by this model, where the flux density is transferred from the luminosity by Eq.~\eqref{eq:Lmin}.
The cyan and lime star markers in the figure represent the estimated 230 GHz flux densities for M87* and 0402+379 by our model. M87* with $z=0.004283$, $M=6.5\times10^9\ M_\odot$ is estimated to be $\sim$1 Jy, consistent with observation \citep{EHTC2019ApJL875.L1}. 0402+379 with $z=0.055$, $M=1.5\times10^{10}\ M_\odot$ is estimated to be $\sim$34 mJy. This is reasonable compared with the 22 GHz flux density observed by \cite{Bansal2017ApJ843.14}.

\subsection{radio luminosity function}
\label{sec:RLF}

In this section we redefine the RLF in per log$_{10} L_\nu$, written as 
\begin{equation}
    \tilde{\Phi}(L_\nu,z)=\frac{dN_{\rm AGN}}{d\left(\log_{10}(L_\nu)\right) dV},
\end{equation}
We use the RLF measured by \cite{Butler2019AA625.111} 
\begin{equation}
    \tilde{\Phi}(L_\nu,z)=\frac{\ln(10)}{\ln(10^{0.4})}\frac{\Phi_0^*(1+z)^{K_{\rm D}}}{\left(\dfrac{L_0^*}{L_{\rm 1.4GHz}}\right)^{\tilde{\alpha}}\!\!\!\!+\left(\dfrac{L_0^*}{L_{\rm 1.4GHz}}\right)^{\tilde{\beta}}}.
    \label{eq:RLF}
\end{equation} 
where $\Phi_0^*$, $L_0^*$, $\tilde{\alpha}$, $\tilde{\beta}$ and $K_{\rm D}$ are model parameters. Here we adopt pure density evolution model, which assumes the AGN population is changing in volume density at redshift $z$.

\begin{table*}
    \centering
    \begin{minipage}[b]{0.9\hsize}
    \caption{Model parameters in RLFs}
    \renewcommand{\arraystretch}{1.3}
    \setlength\tabcolsep{0.3cm}
    \begin{tabular}{ c|ccccc}
         \hline\hline\label{tab:RLF}
        AGN types & $\log_{10}\Phi_0^*$ (Mpc$^{-3}$) & $\log_{10}L_0^*$ (W Hz$^{-1}$) & $\tilde{\alpha}$ &  $\tilde{\beta}$& $K_{\rm D}$\\\hline
        LERGs &  -6.535&  25.910&  -0.472&  -1.382& 0.671\\
        HERGs &  -7.805&  26.776&  -0.516&  -2.0& 1.812\\\hline
    \end{tabular}
    \end{minipage}
\end{table*}
\cite{Butler2019AA625.111} estimated the RLFs from 6287 radio sources in the ultimate X-ray Multi-Mirror Mission (XMM) extragalactic survey south (XXL-S) catalogue. The estimates are separate for two types of AGNs: the low-excitation radio galaxies (LERGs) and the high-excitation radio galaxies (HERGs). We calculate $\tilde{\Phi}(L_\nu,z)$ for each AGN type and sum them together. The values of the parameters can be found in Table~\ref{tab:RLF}, which are summarized from the Table 13 and 14 in \cite{Butler2019AA625.111}. Eq.~\eqref{eq:RLF} is defined at
1.4 GHz, we use the spectral index $\alpha$ to transfer it to higher frequency. We also use the relation
\begin{equation}
    \tilde{\Phi}_1(L_\nu,z)\frac{dV_1}{dz}=\tilde{\Phi}_2(L_\nu,z)\frac{dV_2}{dz},
\end{equation}
to covert RLF to the same cosmology as this work (where 1, 2 refer to different cosmology). In the right panel of Fig.~\ref{fig:appendix}, we plot the predicted number of AGNs observed at 230 GHz with flux density larger than $F_{\rm min}$. It is calculated from integrating the RLF, i.e. Eq.~\eqref{eq:RLF}. We also plot the predicted number based on other RLFs (model B in \citealp{Yuan2016ApJ820.65} and Model C in Erratum of \citealp{Yuan2017ApJ846.78}). We can see the good consistency of the predictions by different RLFs in the local universe ($z\leq0.5$). The plot also includes the observed number of AGNs with flux densities greater than $F_{\rm min}$ according to the catalogue of The Atacama Large Millimeter/submillimeter Array (ALMA). We adopt different values of $\alpha$ and find out that when $\alpha=-0.5$, the predicted numbers of bright AGNs ($\gtrsim 500$ mJy) are reasonable compared with the current observations at 230 GHz (the rightmost three cross markers in the right panel of Fig.~\ref{fig:appendix}). 

\section{Results} 
\label{sec:results}

\subsection{PDF of detectable SMBHBs}

\begin{figure*}
    \centering
    \includegraphics[scale=0.8]{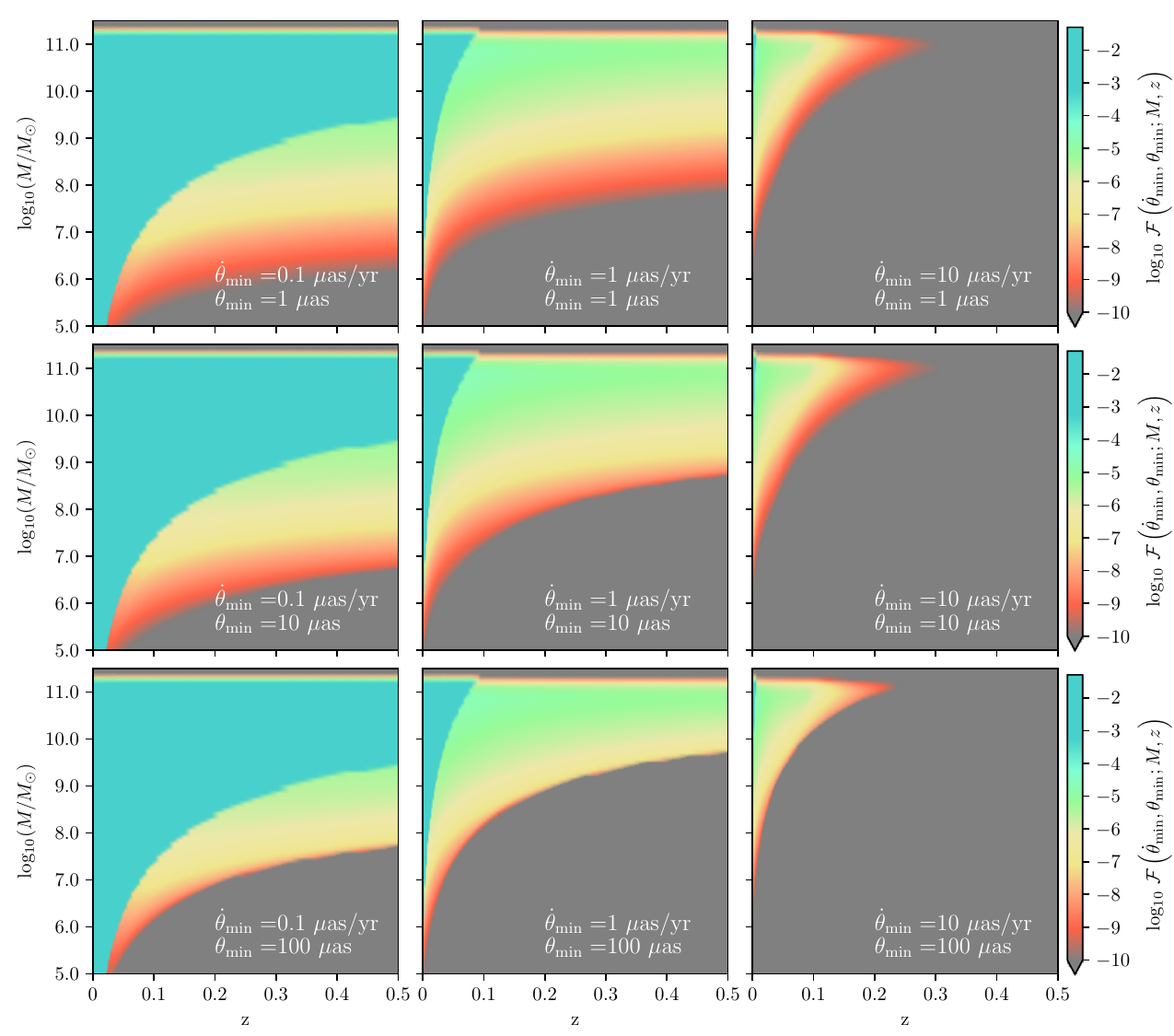}
    \caption{The probability distribution function of a AGN with mass $M$ and redshift $z$ holding a detectable SMBHB, with the minimum detectable angular velocity $\dot{\theta}_{\rm min}$=0.1, 1, 10 $\mu$as/yr and minimum detectable angular separation $\theta_{\rm min}$=1, 10, 100 $\mu$as. Systems with larger $M$ and smaller $z$ have larger PDF. $\dot{\theta}_{\rm min}$ shapes the main structure of the results; $\theta_{\rm min}$ affects the detectability of the low PDF cases.} 
    \label{fig:fraction}
\end{figure*}

We calculate the PDF by Eq.~\eqref{eq:PDF}. Fig.~\ref{fig:fraction} shows the results over $M$ and $z$ for the cases of $\dot{\theta}_{\rm min}=$0.1, 1, 10 $\mu$as/yr, $\theta_{\rm min}$=1, 10, 100 $\mu$as. 
We can see that systems with larger $M$ and smaller $z$ have larger PDF, meaning they are more likely to be detected. In Fig.~\ref{fig:fraction}, the impact of $\dot{\theta}_{\rm min}$ on results (differences among columns) is very different from the impact of $\theta_{\rm min}$ on results (differences among rows). $\dot{\theta}_{\rm min}$ shapes the main structure of the results while $\theta_{\rm min}$ affects the detectability of low PDF cases. Since $\dot{\theta}_{\rm min}$ determines the upper limit of the detectable period $P_{\rm max}$, Eq.~\eqref{eq:Pmax}, while $\theta_{\rm min}$ determines the lower limit $P_{\rm min}$, Eq.~\eqref{eq:Pmin}. It means $\dot{\theta}_{\rm min}$ constrain the detectability of long-period and far-separated systems. In contrast, $\theta_{\rm min}$ limits the detectability of short-period and small-orbit systems. And since the integrated resident timescale in the long period region is much larger than that in the short period region, detectable systems are more likely to occur in long period systems.  To sum up,  $\dot{\theta}_{\rm min}$ plays a more substantial role in constraining high PDF cases.

As shown in Fig.~\ref{fig:fraction}, the blue region has high PDF, which exists in the region with large mass, small redshift and small $\dot{\theta}_{\rm min}$. It corresponds to the scenario that the relative angular velocity of the binary is always larger than $\dot{\theta}_{\rm min}$, so the entire dynamical friction stage can be observed. In this case, $\theta_{\rm min}$ is relatively unimportant. The green-to-red region shows that the PDF decreases from $\sim10^{-4}$ to $\sim 10^{-9}$ with decreasing mass and increasing redshift. This region corresponds to the case where dynamical friction stages cannot be observed, and $\dot{\theta}_{\rm min}$ and $\theta_{\rm min}$ are both crucial to constraining the result. The gray region is where the PDF is so small that it can be approximated as a non-detectable region.  

It should be noted that, the PDF is independent to observing frequency and $F_{\rm min}$, and is only determined by $\dot{\theta}_{\rm min}$ and $\theta_{\rm min}$. 

\subsection{Number of detectable SMBHBs at 86 to 690 GHz}

\begin{table*}[]
    \centering
    \begin{minipage}[b]{0.9\hsize}
    \caption{The number of detectable SMBHBs}
    \renewcommand{\arraystretch}{1.3}
    \setlength\tabcolsep{0.7cm}
    \begin{tabular}{ c|cccc}
         \hline\hline\label{tab:Num}
         $\dot{\theta}_{\rm min}$  & \multicolumn{4}{c}{Num. of detect. SMBHBs$^*$}\\
         ($\mu$as/yr) & 86 GHz & 230 GHz & 345 GHz & 690 GHz\\\hline
         3  & 1 &  0 & 0 & 0\\
         1 & 20 & 18 & 17 & 15\\
         0.1 & 279 & 140 & 105 & 64\\[0.5em] \hline
    \end{tabular}
    \end{minipage}
    \begin{minipage}[b]{0.7\hsize}
    \centering
    \begin{flushleft}
        \ \\
        $^*$ $\theta_{\rm min}=$ 40, 15, 10 and 5 $\mu$as respectively for 86, 230, 345 and 690 GHz; \\
        \textcolor{white}{$^*$} $F_{\rm min}=$ 10 mJy for all cases.
    \end{flushleft}
    \end{minipage}

\end{table*}

\begin{figure*}
    \centering
    \includegraphics[scale=0.8]{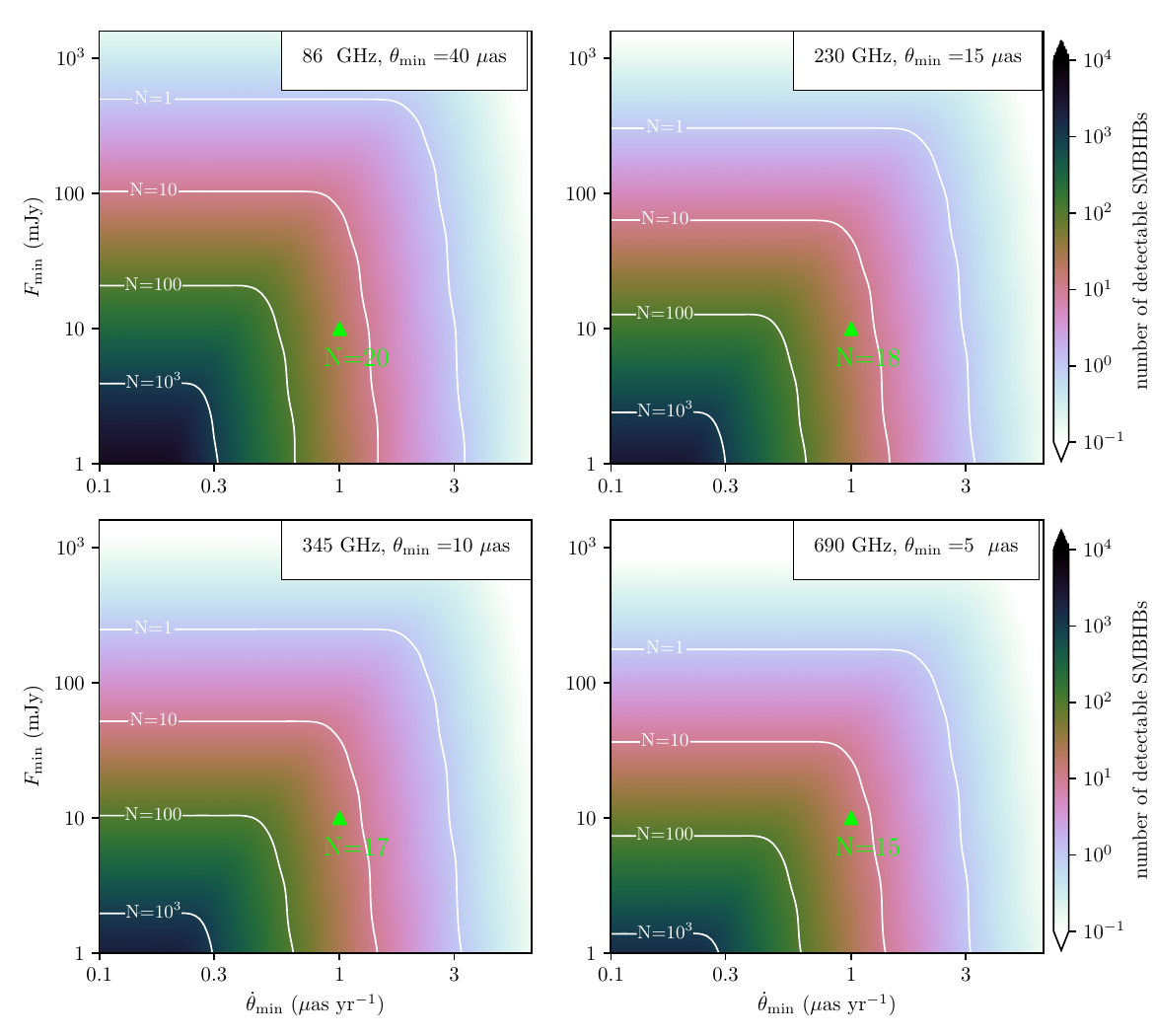}
    \caption{The number of detectable SMBHBs at 86, 230, 345 and 690 GHz, with  $\theta_{\rm min}=40, 15, 10, 5\ \mu$as respectively. The detectable numbers are N=20, 18, 17, 15 at each frequencies under the target capability of sub/millimeter VLBI ($F_{\rm min}=10$ mJy and $\dot{\theta}_{\rm min}=1\ \mu$as/yr, marked with lime triangles). }
    \label{fig:Num_pm_res}
\end{figure*}

We calculate the number of detectable SMBHBs by Eq.~\eqref{eq:N_SMBHBs} for the cases of 86, 230, 345, and 690 GHz. $\theta_{\rm min}$ is fixed as 40, 15, 10, 5 $\mu$as at each frequency, which are the values under the best angular resolution. The results with varying $F_{\rm min}$ and $\dot{\theta}_{\rm min}$ are shown in Fig.~\ref{fig:Num_pm_res}. In Tab.~\ref{tab:Num} we display the results of $\dot{\theta}_{\rm min}$=3, 1, 0.1 $\mu$as/yr at fixed $F_{\rm min}$=10 mJy. We find that when $\dot{\theta}_{\rm min}\sim3\ \mu$as/yr, the 86 GHz VLBI can detect at least one SMBHB. When $\dot{\theta}_{\rm min}\sim1\ \mu$as/yr, the sub/millimeter VLBI can detect $\sim$20 systems (also see Fig.~\ref{fig:Num_pm_res}). If $\dot{\theta}_{\rm min}\sim0.1\ \mu$as/yr, more than a hundred become detectable at 86-345 GHz. When consider the same $F_{\rm min}$ and $\dot{\theta}_{\rm min}$ from 86 GHz to 690 GHz, VLBI at 86 GHz can detect more SMBHBs, because the sources are more brighter at that frequency. The difference between low and high frequency results depends on the spectral index $\alpha$, and a steeper spectra may enlarge the differences. 

\subsection{distributions in the physical parameter space}

\begin{figure*}
    \centering
    \includegraphics[scale=0.8]{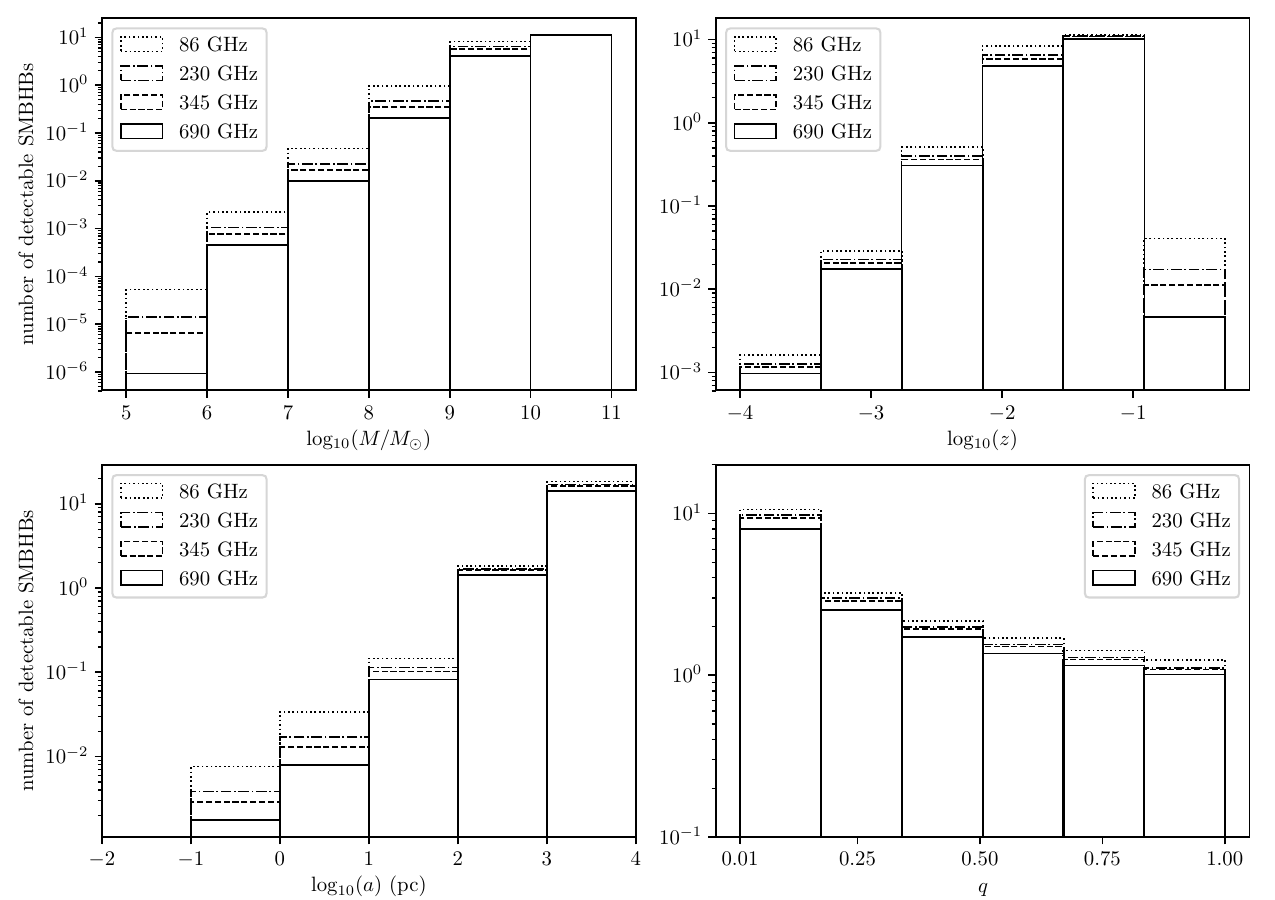}
    \caption{The number distribution of detectable SMBHBs over total mass $M$ (top left), redshift $z$ (top right), separation $a$ (bottom left) and mass ratio $q$ (bottom right). Here we fixed $F_{\rm min}=10$ mJy, $\dot{\theta}_{\rm min}=1\ \mu$as/yr, $\theta_{\rm min}=$40, 15, 10, 5 $\mu$as respectively for 86, 230, 345, 690 GHz. }
    \label{fig:Num_distribution}
\end{figure*}

We additionally analyze the results of $F_{\rm min}=10$ mJy and $\dot{\theta}_{\rm min}=1\ \mu$as/yr at different frequencies and show their distributions in the physical parameter space of $M$, $z$, $a$ and $q$, see Fig.~\ref{fig:Num_distribution}. We notice that the most detectable systems are in the AGN sources with a high mass $10^9$-$10^{11}\ M_\odot$ and a relatively low redshift $z\lesssim 0.1$. The binaries with separations smaller than $\sim$0.1 pc are hardly to be detected. $\sim$0.1-100pc cases are relatively easier, but only be in a small fraction of all the detectable cases. Most detectable binaries are separated by $\sim$100 pc to $\sim$10 kpc. The results are generally insensitive to mass ratio $q$, but the number is slightly increased for smaller $q$. It is because systems with smaller $q$ appear to have a longer lifetime, which gives them a slightly higher chance of being detected.  Finally, the binaries which are detectable at low frequency but undetectable at high frequency are mainly concentrated in the smaller $M$, larger $z$ and smaller $a$.

\section{Towards the observational requirements}
\label{sec:techniques}

On the aspect of VLBI obervation, $\dot{\theta}_{\rm min}$ is determined by proper motion precision, $\theta_{\rm min}$ is determined by angular resolution, $F_{\rm min}$ is determined by baseline sensitivity. The angular resolution is proportional to the observed wavelength over the telescope aperture. Global VLBI can perform like an earth-size telescope, so the angular resolution can reach 40, 15, 10, 5 $\mu$as at 86, 230, 345, 690 GHz, respectively. In this section, we discuss the feasibility of achieving 1 $\mu$as proper motion precision and 10 mJy baseline sensitivity required to detect $\sim$20 SMBHBs.

\subsection{1 $\mu$as yr$^{-1}$ proper motion precision}
Proper motion precision is determined by astrometry accuracy and observational period. Here we focus on the discussion of astrometry accuracy. It depends on a variety of factors, such as source structure, atmosphere, antenna location, instrumentation, and thermal noise. \citep[see e.g.,][]{Reid2014ARAA52.339}. For the ideal case of point sources, we only need to consider atmospheric errors and thermal noise.

VLBI can reach extremely high astrometry accuracy due to its long baselines. But it also introduces difficulties in calibrating the uncorrelated atmospheric errors of the widely distributed arrays. One way to solve the problem is nodding between the target source and the reference source, then use the phase-referencing (PR) method to calibrate the atmospheric errors. However, due to the tropospheric fluctuations, the traditional PR is only applicable for low frequency scenarios \citep[$\lesssim$43 GHz, see, e.g.,][]{Rioja2020AARv28.6}. A new technique called source-frequency phase-referencing (SFPR) can effectively calibrate the tropospheric errors and make sub/millimeter VLBI astrometry achievable \citep{Rioja2011AJ141.114,Zhao2019JKAS52.23,Jiang2021ApJ922.L16}. It requires simultaneous multiple-frequency observation and uses the target source's low frequency observation as a reference to calibrate the high frequency observations. For the idealized situation, by taking the SFPR technique, the astrometry accuracy only depends on the residual ionospheric error and the thermal noise. The ionospheric error is approximated to \citep[e.g., supplementary information of][]{Hada2011Natur477.185}
\begin{equation}
        \delta\theta_{\rm ion}\sim\frac{c\delta\tau_{\rm ion}}{B}\sec{Z}\tan{Z}\delta Z,
\end{equation}
where $\tau_{\rm ion}$ is the residual ionospheric zenith delay which is calculated by $c\delta\tau_{\rm ion}=40.3\ (\delta I/\rm{m^{-2}})(\nu/\rm{Hz})^{-2}$ 
 (m) \citep{Reid2014ARAA52.339}. $\delta I$ is residual total electron content and we adopt the typical value of the sky condition over VLBA stations $\delta I\sim 1.8\times 10^{16}$ m$^{-2}$ \citep[supplementary of][]{Hada2011Natur477.185}. $Z$ is the zenith angle, $\delta Z$ is the angular separation of the two sources, $B$ is the baseline. If we adopt $B=1.07\times10^7$ m (the longest baseline of EHT 2017 array), 230 GHz, $\delta Z=100'$, $Z\lesssim 70^\circ$, then the ionosphere error is $\delta\theta_{\rm ion}\lesssim 0.001\ \mu$as, which is small enough to be neglected. 

The thermal noise can be simply described as \citep[][]{Reid2014ARAA52.339}
\begin{equation}
    \delta\theta_{\rm thermal}\sim 0.5\frac{\theta_{\rm beam}}{\rm SNR}
\end{equation}
where $\theta_{\rm beam}$ is the synthesized beam size and $\rm SNR$ is the image signal-to-noise ratio. For 230 GHz, $\theta_{\rm beam}\sim 15\ \mu$as, if take $\rm SNR\sim30$ and observe one year, the error of proper motion measurement caused by thermal noise is $\dot{\theta}_{\rm thermal}\sim 0.25\ \mu$as/yr. 

\begin{figure*}

    \centering
    \includegraphics[scale=0.8]{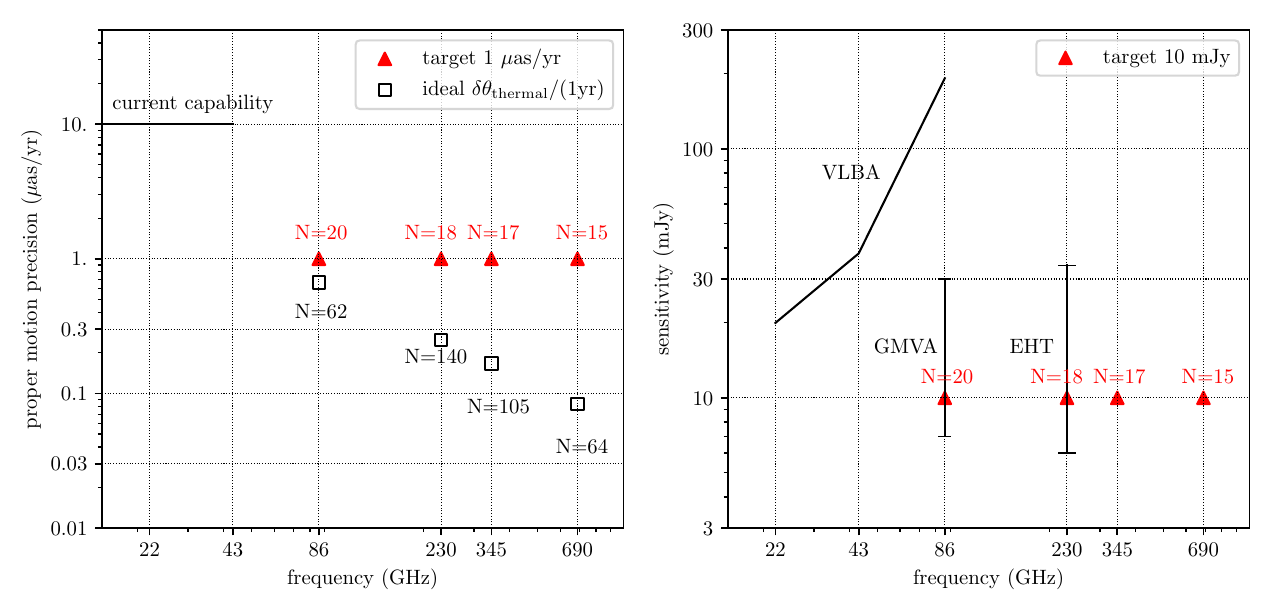}

    \caption{
        Left: The target proper motion precision ($1\ \mu$as/yr, marked with red triangles) required for tracking relative motion of a SMBHB by sub/millimeter VLBI, and the ideal proper motion precision (marked with black squares) with only considering the thermal noise over a one-year observational period, at 86, 230, 345 and 690 GHz. The numbers of detectable SMBHBs for which $\dot{\theta}_{\rm min}$ adopt each case of proper motion precision are also shown. The solid line refers to the current capability of proper motion measurement $\sim 10\ \mu$as/yr at frequency lower than 43 GHz \citep{Rioja2020AARv28.6}. Right: The target sensitivity of sub/millimeter at 86, 230, 345 and 690 GHz ($10$ mJy, marked by red triangles). The numbers of detectable SMBHBs are shown above the markers.  For comparison, we plot the sensitivity of VLBA at 22, 43 and 86 GHz, the sensitivity of GMVA at 86 GHz, and the sensitivity of EHT at 230 GHz, based on Table~\ref{tab:sensit}. }
    \label{fig:submmVLBI_capability}    

\end{figure*}

Both of the ionospheric error and the thermal noise are smaller at higher frequencies, that is one of the advantages of sub/miliimeter VLBI. Current proper motion measurements at observing frequency $\lesssim$ 43 GHz is about 10 $\mu$as/yr \citep{Rioja2020AARv28.6}.  The left panel in Fig.~\ref{fig:submmVLBI_capability} shows the proposed precision of proper motion for sub/milimeter VLBI (1 $\mu$as/yr), the ideal case of only considering the error caused by the thermal noise, and the current capability at low frequencies.

\subsection{ 10 mJy sensitivity}

\begin{table*}[]
\centering
    \begin{minipage}[b]{1\hsize}
    \caption{Baseline sensitivity for 7 $\sigma$ detection threshold\small\textsuperscript{a}}
    \renewcommand{\arraystretch}{1.3}
    \setlength\tabcolsep{0.38cm}
    \begin{tabular}{c|cccccc}
         \hline\hline\label{tab:sensit}
         & frequency & SEFD$_1$  & SEFD$_2$ & bandwidth  & integration time  & sensitivity \\
         & (GHz) & (Jy) & (Jy) & (GHz) & (s) & (mJy)\\\hline
         VLBA\small\textsuperscript{b} & 22 & 640 & 640 & 0.512 & 60 & 20\\
         & 43 & 1181 & 1181 & 0.512 & 60 & 38\\
         & 86 & 4236 & 4236 & 0.512 & 30 & 192\\\hline
         GMVA\small\textsuperscript{c} & 86 & 74.9 & 192 & 0.512 & 20 & 7\\
         & 86 & 74.9 & 3238 & 0.512 & 20 & 30\\\hline
         EHT\small\textsuperscript{d}  & 230 & 74 & 700 & 4 & 10 & 6\\
         & 230 & 74 & 19300 & 4 & 10 & 34\\[0.5em] \hline
    \end{tabular}
    \end{minipage}

    \begin{minipage}[b]{0.85\hsize}
    \centering
    \begin{flushleft}
        \ \\
    \small\textsuperscript{a} $\sigma$ is estimated by Eq.~\eqref{eq:sensit} with $\eta=0.88$ (2-bit samples).\\
    \small\textsuperscript{b} VLBA capabilities are listed in Table 5.1, VLBA Observational Status Summary 2024A (\url{https://science.nrao.edu/facilities/vlba/docs/manuals/oss/referencemanual-all-pages})
    VLBA sensitivity is estimated in the case of two identical antennas. \\
    \small\textsuperscript{c} GMVA capabilities can be found on its website (\url{https://www3.mpifr-bonn.mpg.de/div/vlbi/globalmm/}). We show the highest and lowst baseline sensitivity to ALMA (SEFD=74.9 Jy). The highest sensitivity is for the ALMA-GBT (Green Bank Telescope) baseline , with GBT's SEFD of 192 Jy. The lowest sensitivity is for the ALMA-ON  (Onsala Space Observatory) baseline, with ON's SEFD of 3838 Jy.  \\
    \small\textsuperscript{d} EHT capabilities are obtained from Table 3 in \cite{EHTC2019ApJL875.L2}. We show the sensitivity of ALMA baselines, based on SEFD=74 Jy of ALMA. The highest sensitivity comes from the ALMA-NOEMA baseline, with NOEMA's SEFD of 700 Jy. The lowest sensitivity comes from the ALMA-SPT (South Pole Telescope) baseline, with SPT's SEFD of 19300 Jy.
    \end{flushleft}
    \end{minipage}

\end{table*}

Improving sensitivity of VLBI requires reducing thermal noise of each baseline. Thermal noise is described as a Gaussian-distributed error with the standard deviation \citep{Thompson2017isra.book}
\begin{equation}
    \label{eq:sensit}
    \sigma_{ij}=\frac{1}{\eta}\sqrt{\frac{\rm{SEFD}_i\times\rm{SEFD_j}}{2\Delta\nu\Delta t}}.
\end{equation}
The subscript i, j refer to two stations and $\eta$ is the digital loss in the signal digitization. SEFD is the system equivalent flux density of the station, which is defined as $2k_{\rm B}T_{\rm sys}/A_{\rm eff}$, where $k_{\rm B}$ is the Boltzmann constant, $T_{\rm sys}$ is the system temperature and $A_{\rm eff}$ is the effective area. $\Delta\nu$ is the bandwidth and $\Delta t$ is the integration time. 

To minimize $\sigma_{ij}$, one way is to improve the SEFD of each station. For example, ALMA with $A_{\rm eff}\sim73$ m and $T_{\rm sys}\sim76$ K achieves SEFD $\sim$74 Jy, which boost the EHT sensitivity to $\sigma\lesssim 1$ mJy at 230 GHz on baselines to ALMA\citep{EHTC2019ApJL875.L2}. The second way is to increase the bandwidth. For example, the ngEHT will double the bandwidth to increase its sensitivity \citep{Doeleman2019BAAS51.256}. The third way is to increase the integration time. The integration time is extremely short at high frequency ($\lesssim$ 10 s at 230 GHz). 
 But if use the the frequency phase transfer \citep{Rioja2011AJ141.114,Rioja2023Galax11.16} and further being refined by SFPR method, it can be extended to several hours \citep{galaxies11010003}. The baseline sensitivity for $7\sigma$ detection threshold of some present facilities are: 192 mJy at 86 GHz for the Very Long Baseline Array (VLBA), 7-30 mJy at 86 GHz for the Global mm-VLBI Array (GMVA), and 6-34 mJy at 230 GHz for the EHT. It can be seen in Table~\ref{tab:sensit} for more details. So by taking the new observation schemes, the proposed sensitivity of 10 mJy is reasonably achievable. The right panel in Fig.~\ref{fig:submmVLBI_capability} shows the sensitivity of sub/millimeter VLBI and the present telescopes.

\subsection{simultaneous multi-frequency observation}

Following the above discussions, simultaneous multi-frequency observations are very helpful for observing SMBHBs by sub/milimeter VLBI. The crucial reason is that without such a technique, it is very hard to reach a $\sim1\ \mu$as/yr proper motion precision. In addition, this technique makes arrays more sensitive and allows them to relax their other stringent requirements to reach high sensitivity. It can achieve the required capabilities to observe SMBHBs by applying SFPR. More specifically, we propose two ways to realize such observations:
\begin{enumerate}
    \item simultaneous multi-frequency relative astrometry \\
    For the SMBHB with a small orbital separation, the two black holes could be covered by the primary beam of single antenna dish, e.g. 18\,arcsec ($\sim$18\,kpc at a distance of $z=0.05$) of 15\,meter antenna beam at 230\,GHz. This includes most the detectable SMBHBs in our calculation (Fig.~\ref{fig:Num_distribution}). The ideal case for performing in-beam SFPR on SMBHB observation is that the residual ionospheric error is negligible and the astrometric precision could approach the limit of thermal noise, as shown in Fig.~\ref{fig:submmVLBI_capability}. The visible SMBHB 0402+379 with a orbital separation 7.3\,pc \citep[$\sim$7.3\,mas in the sky, see][]{Bansal2017ApJ843.14} would serve as an good example of the application of such observations. With the help of SFPR to remove the atmospheric errors and increase the coherent integration time \citep{galaxies11010003}, their relative motions can be precisely tracked. 
    \item simultaneous multi-frequency absolute astrometry \\
    If the two sources are too separated to use in-beam observation, the relative astrometry becomes inapplicable. It probably occurs in the case of the nearby binary systems under the early dynamical friction stages, such as a binary system with a few kpc separation at a distance $z\lesssim0.01$. The orbital motions of such far-separate nearby SMBHBs can only be monitored by phase referencing to one or more background calibrators, and it should be performed through absolute astrometry. This puts stringent requirements for sub/millimeter VLBI observations: It is necessary to find at least one in-beam calibrator to track the motions of one black hole in the binary system. If without in-beam calibrator, the co-located antennas or dual-beam system are requested to avoid nodding antenna between calibrators and targets frequently. SFPR could help to relax the sensitivity requirement of finding in-beam calibrators and confirm the motions at higher frequencies simultaneously for absolute astrometry observation.  
\end{enumerate}

\section{conclusion and discussion}
\label{sec:conclusion}
We extend the methodology of \cite{DOrazio2018ApJ863.185}
to link the observable constraints with the orbital evolution of SMBHBs, and estimate the number of detectable SMBHBs through directly tracking their relative motions by VLBI at 86 to 690 GHz.

We find that:
\begin{itemize}
    \item $\sim$ 20 SMBHB systems are expected to be detected at 86 GHz under the capability of $\sim 1\ \mu$as/yr proper motion precision, 10 mJy sensitivity and 40 $\mu$as resolution. 
    \item If the proper motion precision reaches $\sim 0.1\ \mu$as/yr, more than a hundred SMBHBs become detectable at 86-345 GHz.
    \item The detectable systems are concentrated in the parameter space of mass in $\sim10^9-10^{11}\ M_\odot$, redshift $\lesssim0.1$, separation $\gtrsim0.1$ pc. It is relatively insensitive to the mass ratio.  
\end{itemize}
We also show that the simultaneous multi-frequency observation is very helpful in reaching the capabilities of $\sim 1\ \mu$as/yr and $\sim$10 mJy, and propose the ways of relative astrometry and absolute astrometry to realize such detection with the help of SFPR technique.

In this work, we did not consider the brightness distribution of the two sources in a binary system. Simulations show that the smaller black hole in the binaries accretes more mass from the gas environment \citep{Munoz2020ApJ889.114,Lai2022arXiv2211.00028}, which introduces complications in modeling the relation of the brightness ratio and the mass ratio. If one component is much dimer than the other, the sensitivity requirements may become more stringent.   

Our results are based on the assumption of face-on circular orbit. For eccentric binary systems, the evolution of separation and eccentricity are coupled, and such evolution is highly dependent on the environment \citep[e.g.][]{Lai2022arXiv2211.00028}. The estimation of inclination between the system and the observer requires a sufficient long VLBI observation time to cover a large enough part of the binary period \citep[e.g.][]{Fang2022ApJ927.93}. The framework introduced in this work has the potential to be extended to include eccentricity and inclination in future work. 

Since the separation of observable systems is much larger than the scale of the source, the assumptions of point source and source located at the center of black hole are valid. But when the separation reaches the source-scale-order, which is beyond the observational capabilities discussed in this work, the structure of the source and the offset between the centre of the source and the centre of the black hole should be taken into consideration. For instance, the frequency-dependent emission region of the jet may need to be considered \citep[e.g., appendix B in][]{DOrazio2018ApJ863.185}. The gravitational lensing effect caused by the rotating black hole binary may also affect the observed image \citep[e.g.,][]{Bohn2015CQG32.6.065002}. 

Using a simple luminosity model, we predict a linear relationship between mass and luminosity at different redshift. In terms of its limitations, this model appears to be best suited for predicting M87-like AGNs, while showing a poor ability to predict Sgr~A*-like sources. It is possible to overcome such limitations by introducing the accretion rate. For instance, in \cite{Pesce2021ApJ923.260}, an Eddington ratio distribution function is used to predict an AGN's mass accretion rate. The luminosity at a particular frequency is then calculated using a model representing the spectral energy distribution, in which synchrotron emission, inverse Compton emission, and bremsstrahlung emission are all taken into account. The use of a more physical luminosity model will be a critical step for future work.

Finally, the results are very weakly related to the galaxy size. The initial separation $a_{\rm ini}$ in this analysis is set to 10 kpc, which is the Milky Way radius scale. The maximum impact parameter $b_{\rm max}$ in Coulomb logarithm, i.e., Eq.~\eqref{eq:Lambda}, is also chosen as 10 kpc. Since the Milky Way is a medium-mass galaxy, this value is reasonable. If we set $a_{\rm ini}$ and $b_{\rm max}$ to the M87-like massive galaxy radius, 40 kpc \citep{Cohen1997ApJ486.230}, then the number of detectable SMBHBs increases by less than 0.7\%. So galaxy size has a negligible impact on the our results.

\begin{acknowledgments}
We thank the reviewer's very detailed and helpful comments. This work is computationally supported by the computing cluster of Shanghai VLBI correlator, which is supported by the Special Fund for Astronomy from the Ministry of Finance in China. This work is financially supported by the China Postdoctoral Science Foundation fellowship (2020M671266); Shanghai post-doctoral excellence Program (2020481); Max Planck Partner Group of the MPG and the CAS; the Key Program of the National Natural Science Foundation of China (grant No. 11933007); the National Natural Science Foundation of China (grant Nos. 12173074, 11803071); the Key Research Program of Frontier Sciences, CAS (grant No. ZDBS-LY-SLH011, QYZDJ-SSW-SLH057) and the Shanghai Pilot Program for Basic Research – Chinese Academy of Science, Shanghai Branch (JCYJ-SHFY-2022-013).
\end{acknowledgments}

%

\vspace{5mm}







\bibliography{ref}{}

\begin{thebibliography}{}
\expandafter\ifx\csname natexlab\endcsname\relax\def\natexlab#1{#1}\fi
\providecommand{\url}[1]{\href{#1}{#1}}
\providecommand{\dodoi}[1]{doi:~\href{http://doi.org/#1}{\nolinkurl{#1}}}
\providecommand{\doeprint}[1]{\href{http://ascl.net/#1}{\nolinkurl{http://ascl.net/#1}}}
\providecommand{\doarXiv}[1]{\href{https://arxiv.org/abs/#1}{\nolinkurl{https://arxiv.org/abs/#1}}}

\bibitem[{{An} {et~al.}(2018){An}, {Mohan}, \& {Frey}}]{An2018RaSc53.1211}
{An}, T., {Mohan}, P., \& {Frey}, S. 2018, Radio Science, 53, 1211,
  \dodoi{10.1029/2018RS006647}

\bibitem[{{Arzoumanian} {et~al.}(2021){Arzoumanian}, {Baker}, {Brazier},
  {Brook}, {Burke-Spolaor}, {Becsy}, {Charisi}, {Chatterjee}, {Cordes},
  {Cornish}, {Crawford}, {Cromartie}, {Decesar}, {Demorest}, {Dolch},
  {Elliott}, {Ellis}, {Ferrara}, {Fonseca}, {Garver-Daniels}, {Gentile},
  {Good}, {Hazboun}, {Islo}, {Jennings}, {Jones}, {Kaiser}, {Kaplan}, {Kelley},
  {Key}, {Lam}, {Lazio}, {Luo}, {Lynch}, {Ma}, {Madison}, {McLaughlin},
  {Mingarelli}, {Ng}, {Nice}, {Pennucci}, {Pol}, {Ransom}, {Ray},
  {Shapiro-Albert}, {Siemens}, {Simon}, {Spiewak}, {Stairs}, {Stinebring},
  {Stovall}, {Swiggum}, {Taylor}, {Vallisneri}, {Vigeland}, {Witt}, \&
  {Nanograv Collaboration}}]{Arzoumanian2021ApJ914.121}
{Arzoumanian}, Z., {Baker}, P.~T., {Brazier}, A., {et~al.} 2021, \apj, 914,
  121, \dodoi{10.3847/1538-4357/abfcd3}

\bibitem[{{Bansal} {et~al.}(2017){Bansal}, {Taylor}, {Peck}, {Zavala}, \&
  {Romani}}]{Bansal2017ApJ843.14}
{Bansal}, K., {Taylor}, G.~B., {Peck}, A.~B., {Zavala}, R.~T., \& {Romani},
  R.~W. 2017, \apj, 843, 14, \dodoi{10.3847/1538-4357/aa74e1}

\bibitem[{{Begelman} {et~al.}(1980){Begelman}, {Blandford}, \&
  {Rees}}]{Begelman1980Natur287.307}
{Begelman}, M.~C., {Blandford}, R.~D., \& {Rees}, M.~J. 1980, \nat, 287, 307,
  \dodoi{10.1038/287307a0}

\bibitem[{{Binney} \& {Tremaine}(2008)}]{Binney2008GalacticDynamics}
{Binney}, J., \& {Tremaine}, S. 2008, {Galactic Dynamics: Second Edition}

\bibitem[{{Bohn} {et~al.}(2015){Bohn}, {Throwe}, {H{\'e}bert}, {Henriksson},
  {Bunandar}, {Scheel}, \& {Taylor}}]{Bohn2015CQG32.6.065002}
{Bohn}, A., {Throwe}, W., {H{\'e}bert}, F., {et~al.} 2015, Classical and
  Quantum Gravity, 32, 065002, \dodoi{10.1088/0264-9381/32/6/065002}

\bibitem[{{Breiding} {et~al.}(2021){Breiding}, {Burke-Spolaor}, {Eracleous},
  {Bogdanovi{\'c}}, {Lazio}, {Runnoe}, \& {Sigurdsson}}]{Breiding2021ApJ914.37}
{Breiding}, P., {Burke-Spolaor}, S., {Eracleous}, M., {et~al.} 2021, \apj, 914,
  37, \dodoi{10.3847/1538-4357/abfa9a}

\bibitem[{{Burke-Spolaor} {et~al.}(2018){Burke-Spolaor}, {Blecha},
  {Bogdanovi{\'c}}, {Comerford}, {Lazio}, {Liu}, {Maccarone}, {Pesce}, {Shen},
  \& {Taylor}}]{Burke-Spolaor2018ASPC517.677}
{Burke-Spolaor}, S., {Blecha}, L., {Bogdanovi{\'c}}, T., {et~al.} 2018, in
  Astronomical Society of the Pacific Conference Series, Vol. 517, Science with
  a Next Generation Very Large Array, ed. E.~{Murphy}, 677

\bibitem[{{Butler} {et~al.}(2019){Butler}, {Huynh}, {Kapi{\'n}ska},
  {Delvecchio}, {Smol{\v{c}}i{\'c}}, {Chiappetti}, {Koulouridis}, \&
  {Pierre}}]{Butler2019AA625.111}
{Butler}, A., {Huynh}, M., {Kapi{\'n}ska}, A., {et~al.} 2019, \aap, 625, A111,
  \dodoi{10.1051/0004-6361/201834581}

\bibitem[{{Chandrasekhar}(1943)}]{Chandrasekhar1943ApJ97.255}
{Chandrasekhar}, S. 1943, \apj, 97, 255, \dodoi{10.1086/144517}

\bibitem[{{Chen} {et~al.}(2022{\natexlab{a}}){Chen}, {Ni}, {Tremmel}, {Di
  Matteo}, {Bird}, {DeGraf}, \& {Feng}}]{Chen2022MNRAS510.531}
{Chen}, N., {Ni}, Y., {Tremmel}, M., {et~al.} 2022{\natexlab{a}}, \mnras, 510,
  531, \dodoi{10.1093/mnras/stab3411}

\bibitem[{{Chen} {et~al.}(2022{\natexlab{b}}){Chen}, {Ni}, {Holgado}, {Matteo},
  {Tremmel}, {DeGraf}, {Bird}, {Croft}, \& {Feng}}]{Chen2022MNRAS514.2220}
{Chen}, N., {Ni}, Y., {Holgado}, A.~M., {et~al.} 2022{\natexlab{b}}, \mnras,
  514, 2220, \dodoi{10.1093/mnras/stac1432}

\bibitem[{{Cohen} \& {Ryzhov}(1997)}]{Cohen1997ApJ486.230}
{Cohen}, J.~G., \& {Ryzhov}, A. 1997, \apj, 486, 230, \dodoi{10.1086/304518}

\bibitem[{{Dittmann} \& {Ryan}(2022)}]{Dittmann2022MNRAS513.6158}
{Dittmann}, A.~J., \& {Ryan}, G. 2022, \mnras, 513, 6158,
  \dodoi{10.1093/mnras/stac935}

\bibitem[{{Doeleman} {et~al.}(2019){Doeleman}, {Blackburn}, {Doeleman},
  {Dexter}, {Gomez}, {Johnson}, {Palumbo}, {Weintroub}, {Farah}, {Fish},
  {Loinard}, {Lonsdale}, {Narayanan}, {Patel}, {Pesce}, {Raymond}, {Tilanus},
  {Wielgus}, {Akiyama}, {Bower}, {Broderick}, {Deane}, {Fromm}, {Gammie},
  {Gold}, {Janssen}, {Kawashima}, {Krichbaum}, {Marrone}, {Matthews}, {Mizuno},
  {Rezzolla}, {Roelofs}, {Ros}, {Savolainen}, {Yuan}, {Zhao}, {Blackburn},
  {Doeleman}, {Dexter}, {Gomez}, {Johnson}, {Palumbo}, {Weintroub}, {Farah},
  {Fish}, {Loinard}, {Lonsdale}, {Narayanan}, {Patel}, {Pesce}, {Raymond},
  {Tilanus}, {Wielgus}, {Akiyama}, {Bower}, {Broderick}, {Deane}, {Fromm},
  {Gammie}, {Gold}, {Janssen}, {Kawashima}, {Krichbaum}, {Marrone}, {Matthews},
  {Mizuno}, {Rezzolla}, {Roelofs}, {Ros}, {Savolainen}, {Yuan}, \&
  {Zhao}}]{Doeleman2019BAAS51.256}
{Doeleman}, S., {Blackburn}, L., {Doeleman}, S., {et~al.} 2019, in Bulletin of
  the American Astronomical Society, Vol.~51, 256

\bibitem[{{D'Orazio} \& {Loeb}(2018)}]{DOrazio2018ApJ863.185}
{D'Orazio}, D.~J., \& {Loeb}, A. 2018, \apj, 863, 185,
  \dodoi{10.3847/1538-4357/aad413}

\bibitem[{{Dosopoulou} \& {Antonini}(2017)}]{Dosopoulou2017ApJ840.31}
{Dosopoulou}, F., \& {Antonini}, F. 2017, \apj, 840, 31,
  \dodoi{10.3847/1538-4357/aa6b58}

\bibitem[{{Event Horizon Telescope Collaboration}
  {et~al.}(2019{\natexlab{a}}){Event Horizon Telescope Collaboration},
  {Akiyama}, {Alberdi}, {Alef}, {Asada}, {Azulay}, {Baczko}, {Ball},
  {Balokovi{\'c}}, {Barrett}, \& et~al.}]{EHTC2019ApJL875.L1}
{Event Horizon Telescope Collaboration}, {Akiyama}, K., {Alberdi}, A., {et~al.}
  2019{\natexlab{a}}, The Astrophysical Journal Letters, 875, L1,
  \dodoi{10.3847/2041-8213/ab0ec7}

\bibitem[{{Event Horizon Telescope Collaboration}
  {et~al.}(2019{\natexlab{b}}){Event Horizon Telescope Collaboration},
  {Akiyama}, {Alberdi}, {Alef}, {Asada}, {Azulay}, {Baczko}, {Ball},
  {Balokovi{\'c}}, {Barrett}, \& et~al.}]{EHTC2019ApJL875.L2}
---. 2019{\natexlab{b}}, The Astrophysical Journal Letters, 875, L2,
  \dodoi{10.3847/2041-8213/ab0c96}

\bibitem[{{Event Horizon Telescope Collaboration} {et~al.}(2022){Event Horizon
  Telescope Collaboration}, {Akiyama}, {Alberdi}, {Alef}, {Algaba}, {Anantua},
  {Asada}, {Azulay}, {Bach}, {Baczko}, {Ball}, {Balokovi{\'c}}, {Barrett},
  {Baub{\"o}ck}, {Benson}, {Bintley}, {Blackburn}, {Blundell}, {Bouman},
  {Bower}, {Boyce}, {Bremer}, {Brinkerink}, {Brissenden}, {Britzen},
  {Broderick}, {Broguiere}, {Bronzwaer}, {Bustamante}, {Byun}, {Carlstrom},
  {Ceccobello}, {Chael}, {Chan}, {Chatterjee}, {Chatterjee}, {Chen}, {Chen},
  {Cheng}, {Cho}, {Christian}, {Conroy}, {Conway}, {Cordes}, {Crawford},
  {Crew}, {Cruz-Osorio}, {Cui}, {Davelaar}, {Laurentis}, {Deane}, {Dempsey},
  {Desvignes}, {Dexter}, {Dhruv}, {Doeleman}, {Dougal}, {Dzib}, {Eatough},
  {Emami}, {Falcke}, {Farah}, {Fish}, {Fomalont}, {Ford}, {Fraga-Encinas},
  {Freeman}, {Friberg}, {Fromm}, {Fuentes}, {Galison}, {Gammie}, {Garc{\'\i}a},
  {Gentaz}, {Georgiev}, {Goddi}, {Gold}, {G{\'o}mez-Ruiz}, {G{\'o}mez}, {Gu},
  {Gurwell}, {Hada}, {Haggard}, {Haworth}, {Hecht}, {Hesper}, {Heumann}, {Ho},
  {Ho}, {Honma}, {Huang}, {Huang}, {Hughes}, {Ikeda}, {Impellizzeri}, {Inoue},
  {Issaoun}, {James}, {Jannuzi}, {Janssen}, {Jeter}, {Jiang},
  {Jim{\'e}nez-Rosales}, {Johnson}, {Jorstad}, {Joshi}, {Jung}, {Karami},
  {Karuppusamy}, {Kawashima}, {Keating}, {Kettenis}, {Kim}, {Kim}, {Kim},
  {Kim}, {Kino}, {Koay}, {Kocherlakota}, {Kofuji}, {Koch}, {Koyama}, {Kramer},
  {Kramer}, {Krichbaum}, {Kuo}, {Bella}, {Lauer}, {Lee}, {Lee}, {Leung},
  {Levis}, {Li}, {Lico}, {Lindahl}, {Lindqvist}, {Lisakov}, {Liu}, {Liu},
  {Liuzzo}, {Lo}, {Lobanov}, {Loinard}, {Lonsdale}, {Lu}, {Mao}, {Marchili},
  {Markoff}, {Marrone}, {Marscher}, {Mart{\'\i}-Vidal}, {Matsushita},
  {Matthews}, {Medeiros}, {Menten}, {Michalik}, {Mizuno}, {Mizuno}, {Moran},
  {Moriyama}, {Moscibrodzka}, {M{\"u}ller}, {Mus}, {Musoke}, {Myserlis},
  {Nadolski}, {Nagai}, {Nagar}, {Nakamura}, {Narayan}, {Narayanan},
  {Natarajan}, {Nathanail}, {Fuentes}, {Neilsen}, {Neri}, {Ni}, {Noutsos},
  {Nowak}, {Oh}, {Okino}, {Olivares}, {Ortiz-Le{\'o}n}, {Oyama}, {{\"O}zel},
  {Palumbo}, {Paraschos}, {Park}, {Parsons}, {Patel}, {Pen}, {Pesce},
  {Pi{\'e}tu}, {Plambeck}, {PopStefanija}, {Porth}, {P{\"o}tzl}, {Prather},
  {Preciado-L{\'o}pez}, {Psaltis}, {Pu}, {Ramakrishnan}, {Rao}, {Rawlings},
  {Raymond}, {Rezzolla}, {Ricarte}, {Ripperda}, {Roelofs}, {Rogers}, {Ros},
  {Romero-Ca{\~n}izales}, {Roshanineshat}, {Rottmann}, {Roy}, {Ruiz},
  {Ruszczyk}, {Rygl}, {S{\'a}nchez}, {S{\'a}nchez-Arg{\"u}elles},
  {S{\'a}nchez-Portal}, {Sasada}, {Satapathy}, {Savolainen}, {Schloerb},
  {Schonfeld}, {Schuster}, {Shao}, {Shen}, {Small}, {Sohn}, {SooHoo},
  {Souccar}, {Sun}, {Tazaki}, {Tetarenko}, {Tiede}, {Tilanus}, {Titus},
  {Torne}, {Traianou}, {Trent}, {Trippe}, {Turk}, {van Bemmel}, {van
  Langevelde}, {van Rossum}, {Vos}, {Wagner}, {Ward-Thompson}, {Wardle},
  {Weintroub}, {Wex}, {Wharton}, {Wielgus}, {Wiik}, {Witzel}, {Wondrak},
  {Wong}, {Wu}, {Yamaguchi}, {Yoon}, {Young}, {Young}, {Younsi}, {Yuan},
  {Yuan}, {Zensus}, {Zhang}, {Zhao}, {Zhao}, {Agurto}, {Allardi}, {Amestica},
  {Araneda}, {Arriagada}, {Berghuis}, {Bertarini}, {Berthold}, {Blanchard},
  {Brown}, {C{\'a}rdenas}, {Cantzler}, {Caro}, {Castillo-Dom{\'\i}nguez},
  {Chan}, {Chang}, {Chang}, {Chang}, {Chang}, {Chen}, {Chilson}, {Chuter},
  {Ciechanowicz}, {Colin-Beltran}, {Coulson}, {Crowley}, {Degenaar},
  {Dornbusch}, {Dur{\'a}n}, {Everett}, {Faber}, {Forster}, {Fuchs}, {Gale},
  {Geertsema}, {Gonz{\'a}lez}, {Graham}, {Gueth}, {Halverson}, {Han}, {Han},
  {Hasegawa}, {Hern{\'a}ndez-Rebollar}, {Herrera}, {Herrero-Illana},
  {Heyminck}, {Hirota}, {Hoge}, {Hostler Schimpf}, {Howie}, {Huang}, {Jiang},
  {Jinchi}, {John}, {Kimura}, {Klein}, {Kubo}, {Kuroda}, {Kwon}, {Lacasse},
  {Laing}, {Leitch}, {Li}, {Liu}, {Liu}, {Lin}, {Lu}, {Mac-Auliffe},
  {Martin-Cocher}, {Matulonis}, {Maute}, {Messias}, {Meyer-Zhao},
  {Monta{\~n}a}, {Montenegro-Montes}, {Montgomerie}, {Moreno Nolasco},
  {Muders}, {Nishioka}, {Norton}, {Nystrom}, {Ogawa}, {Olivares}, {Oshiro},
  {P{\'e}rez-Beaupuits}, {Parra}, {Phillips}, {Poirier}, {Pradel}, {Qiu},
  {Raffin}, {Rahlin}, {Ram{\'\i}rez}, {Ressler}, {Reynolds},
  {Rodr{\'\i}guez-Montoya}, {Saez-Madain}, {Santana}, {Shaw}, {Shirkey},
  {Silva}, {Snow}, {Sousa}, {Sridharan}, {Stahm}, {Stark}, {Test},
  {Torstensson}, {Venegas}, {Walther}, {Wei}, {White}, {Wieching}, {Wijnands},
  {Wouterloot}, {Yu}, {Yu (于威)}, \& {Zeballos}}]{EHTC2022ApJ930L12}
---. 2022, \apjl, 930, L12, \dodoi{10.3847/2041-8213/ac6674}

\bibitem[{{Fang} \& {Yang}(2022)}]{Fang2022ApJ927.93}
{Fang}, Y., \& {Yang}, H. 2022, \apj, 927, 93, \dodoi{10.3847/1538-4357/ac4bd7}

\bibitem[{{Hada} {et~al.}(2011){Hada}, {Doi}, {Kino}, {Nagai}, {Hagiwara}, \&
  {Kawaguchi}}]{Hada2011Natur477.185}
{Hada}, K., {Doi}, A., {Kino}, M., {et~al.} 2011, \nat, 477, 185,
  \dodoi{10.1038/nature10387}

\bibitem[{{Jiang} {et~al.}(2021){Jiang}, {Shen}, {Mart{\'\i}-Vidal}, {Wang},
  {Jiang}, \& {Kawaguchi}}]{Jiang2021ApJ922.L16}
{Jiang}, W., {Shen}, Z., {Mart{\'\i}-Vidal}, I., {et~al.} 2021, \apjl, 922,
  L16, \dodoi{10.3847/2041-8213/ac375c}

\bibitem[{Jiang {et~al.}(2023)Jiang, Zhao, Shen, Rioja, Dodson, Cho, Zhao,
  Eubanks, \& Lu}]{galaxies11010003}
Jiang, W., Zhao, G.-Y., Shen, Z.-Q., {et~al.} 2023, Galaxies, 11,
  \dodoi{10.3390/galaxies11010003}

\bibitem[{{Kelley} {et~al.}(2017){Kelley}, {Blecha}, \&
  {Hernquist}}]{Kelley2017MNRAS464.3131}
{Kelley}, L.~Z., {Blecha}, L., \& {Hernquist}, L. 2017, \mnras, 464, 3131,
  \dodoi{10.1093/mnras/stw2452}

\bibitem[{{Kelley} {et~al.}(2019){Kelley}, {Haiman}, {Sesana}, \&
  {Hernquist}}]{Kelley2019MNRAS485.1579}
{Kelley}, L.~Z., {Haiman}, Z., {Sesana}, A., \& {Hernquist}, L. 2019, \mnras,
  485, 1579, \dodoi{10.1093/mnras/stz150}

\bibitem[{{Lai} \& {Mu{\~n}oz}(2023)}]{Lai2022arXiv2211.00028}
{Lai}, D., \& {Mu{\~n}oz}, D.~J. 2023, \araa, 61, 517,
  \dodoi{10.1146/annurev-astro-052622-022933}

\bibitem[{{Merritt}(2013)}]{Merritt2013CQGra30.244005}
{Merritt}, D. 2013, Classical and Quantum Gravity, 30, 244005,
  \dodoi{10.1088/0264-9381/30/24/244005}

\bibitem[{{Merritt} \& {Milosavljevi{\'c}}(2005)}]{Merritt2005LRR8.8}
{Merritt}, D., \& {Milosavljevi{\'c}}, M. 2005, Living Reviews in Relativity,
  8, 8, \dodoi{10.12942/lrr-2005-8}

\bibitem[{{Mingarelli} {et~al.}(2017){Mingarelli}, {Lazio}, {Sesana}, {Greene},
  {Ellis}, {Ma}, {Croft}, {Burke-Spolaor}, \&
  {Taylor}}]{Mingarelli2017NatAs1.886}
{Mingarelli}, C. M.~F., {Lazio}, T. J.~W., {Sesana}, A., {et~al.} 2017, Nature
  Astronomy, 1, 886, \dodoi{10.1038/s41550-017-0299-6}

\bibitem[{{Mu{\~n}oz} {et~al.}(2020){Mu{\~n}oz}, {Lai}, {Kratter}, \&
  {Miranda}}]{Munoz2020ApJ889.114}
{Mu{\~n}oz}, D.~J., {Lai}, D., {Kratter}, K., \& {Miranda}, R. 2020, \apj, 889,
  114, \dodoi{10.3847/1538-4357/ab5d33}

\bibitem[{{Pesce} {et~al.}(2021){Pesce}, {Palumbo}, {Narayan}, {Blackburn},
  {Doeleman}, {Johnson}, {Ma}, {Nagar}, {Natarajan}, \&
  {Ricarte}}]{Pesce2021ApJ923.260}
{Pesce}, D.~W., {Palumbo}, D. C.~M., {Narayan}, R., {et~al.} 2021, \apj, 923,
  260, \dodoi{10.3847/1538-4357/ac2eb5}

\bibitem[{{Peters}(1964)}]{Peters1964PhRv136.1224}
{Peters}, P.~C. 1964, Physical Review, 136, 1224,
  \dodoi{10.1103/PhysRev.136.B1224}

\bibitem[{{Plotkin} {et~al.}(2012){Plotkin}, {Markoff}, {Kelly}, {K{\"o}rding},
  \& {Anderson}}]{Plotkin2012MNRAS419.267}
{Plotkin}, R.~M., {Markoff}, S., {Kelly}, B.~C., {K{\"o}rding}, E., \&
  {Anderson}, S.~F. 2012, \mnras, 419, 267,
  \dodoi{10.1111/j.1365-2966.2011.19689.x}

\bibitem[{{Quinlan}(1996)}]{Quinlan1996NewA1.35}
{Quinlan}, G.~D. 1996, \na, 1, 35, \dodoi{10.1016/S1384-1076(96)00003-6}

\bibitem[{{Reid} \& {Honma}(2014)}]{Reid2014ARAA52.339}
{Reid}, M.~J., \& {Honma}, M. 2014, \araa, 52, 339,
  \dodoi{10.1146/annurev-astro-081913-040006}

\bibitem[{{Rioja} \& {Dodson}(2011)}]{Rioja2011AJ141.114}
{Rioja}, M., \& {Dodson}, R. 2011, \aj, 141, 114,
  \dodoi{10.1088/0004-6256/141/4/114}

\bibitem[{{Rioja} \& {Dodson}(2020)}]{Rioja2020AARv28.6}
{Rioja}, M.~J., \& {Dodson}, R. 2020, \aapr, 28, 6,
  \dodoi{10.1007/s00159-020-00126-z}

\bibitem[{{Rioja} {et~al.}(2023){Rioja}, {Dodson}, \&
  {Asaki}}]{Rioja2023Galax11.16}
{Rioja}, M.~J., {Dodson}, R., \& {Asaki}, Y. 2023, Galaxies, 11, 16,
  \dodoi{10.3390/galaxies11010016}

\bibitem[{{Saade} {et~al.}(2020){Saade}, {Stern}, {Brightman}, {Haiman},
  {Djorgovski}, {D'Orazio}, {Ford}, {Graham}, {Jun}, {Kraft}, {McKernan},
  {Vikhlinin}, \& {Walton}}]{Saade2020ApJ900.148}
{Saade}, M.~L., {Stern}, D., {Brightman}, M., {et~al.} 2020, \apj, 900, 148,
  \dodoi{10.3847/1538-4357/abad31}

\bibitem[{{Sudou} {et~al.}(2003){Sudou}, {Iguchi}, {Murata}, \&
  {Taniguchi}}]{Sudou2003Sci300.1263}
{Sudou}, H., {Iguchi}, S., {Murata}, Y., \& {Taniguchi}, Y. 2003, Science, 300,
  1263, \dodoi{10.1126/science.1082817}

\bibitem[{{Thompson} {et~al.}(2017){Thompson}, {Moran}, \&
  {Swenson}}]{Thompson2017isra.book}
{Thompson}, A.~R., {Moran}, J.~M., \& {Swenson}, George~W., J. 2017,
  {Interferometry and Synthesis in Radio Astronomy, 3rd Edition},
  \dodoi{10.1007/978-3-319-44431-4}

\bibitem[{{Tremaine} {et~al.}(2002){Tremaine}, {Gebhardt}, {Bender}, {Bower},
  {Dressler}, {Faber}, {Filippenko}, {Green}, {Grillmair}, {Ho}, {Kormendy},
  {Lauer}, {Magorrian}, {Pinkney}, \& {Richstone}}]{Tremaine2002ApJ574.740}
{Tremaine}, S., {Gebhardt}, K., {Bender}, R., {et~al.} 2002, \apj, 574, 740,
  \dodoi{10.1086/341002}

\bibitem[{{Vasiliev} {et~al.}(2015){Vasiliev}, {Antonini}, \&
  {Merritt}}]{Vasiliev2015ApJ810.49}
{Vasiliev}, E., {Antonini}, F., \& {Merritt}, D. 2015, \apj, 810, 49,
  \dodoi{10.1088/0004-637X/810/1/49}

\bibitem[{{Volonteri} {et~al.}(2022){Volonteri}, {Pfister}, {Beckmann},
  {Dotti}, {Dubois}, {Massonneau}, {Musoke}, \&
  {Tremmel}}]{Volonteri2022MNRAS514.640}
{Volonteri}, M., {Pfister}, H., {Beckmann}, R., {et~al.} 2022, \mnras, 514,
  640, \dodoi{10.1093/mnras/stac1217}

\bibitem[{{Yuan} {et~al.}(2016){Yuan}, {Wang}, {Zhou}, \&
  {Mao}}]{Yuan2016ApJ820.65}
{Yuan}, Z., {Wang}, J., {Zhou}, M., \& {Mao}, J. 2016, \apj, 820, 65,
  \dodoi{10.3847/0004-637X/820/1/65}

\bibitem[{{Yuan} {et~al.}(2017){Yuan}, {Wang}, {Zhou}, {Qin}, \&
  {Mao}}]{Yuan2017ApJ846.78}
{Yuan}, Z., {Wang}, J., {Zhou}, M., {Qin}, L., \& {Mao}, J. 2017, \apj, 846,
  78, \dodoi{10.3847/1538-4357/aa8463}

\bibitem[{{Zhao} {et~al.}(2019){Zhao}, {Jung}, {Sohn}, {Kino}, {Honma},
  {Dodson}, {Rioja}, {Han}, {Shibata}, {Byun}, {Akiyama}, {Algaba}, {An},
  {Cheng}, {Cho}, {Cui}, {Hada}, {Hodgson}, {Jiang}, {Lee}, {Lee}, {Niinuma},
  {Park}, {Ro}, {Sawada-Satoh}, {Shen}, {Tazaki}, {Trippe}, {Wajima}, \&
  {Zhang}}]{Zhao2019JKAS52.23}
{Zhao}, G.-Y., {Jung}, T., {Sohn}, B.~W., {et~al.} 2019, Journal of Korean
  Astronomical Society, 52, 23, \dodoi{10.5303/JKAS.2019.52.1.23}

\end{thebibliography}
\bibliographystyle{aasjournal}



\end{document}